\documentclass[aps,prd,reprint,preprintnumbers,superscriptaddress,nofootinbib,floatfix,longbibliography]{revtex4-1}
\usepackage{graphicx,bm,amsmath,amsfonts,amssymb,epstopdf,natbib,hyperref,color,verbatim,multirow,bm,tikz,xcolor}
\enlargethispage{\baselineskip}
\hypersetup{colorlinks=true,urlcolor=blue,citecolor=blue,linkcolor=blue,menucolor=blue,anchorcolor=blue,filecolor=blue}
\everymath{\displaystyle}
\usepackage[normalem]{ulem}
\definecolor{lime}{HTML}{A6CE39}
\DeclareRobustCommand{\orcidicon}{
	\begin{tikzpicture}
	\draw[lime, fill=lime] (0,0) 
	circle [radius=0.16] 
	node[white] {{\fontfamily{qag}\selectfont \tiny ID}};
	\draw[white, fill=white] (-0.0625,0.095) 
	circle [radius=0.007];
	\end{tikzpicture}
	\hspace{-3mm}
}
\foreach \x in {A, ..., Z}{\expandafter\xdef\csname orcid\x\endcsname{\noexpand\href{https://orcid.org/\csname orcidauthor\x\endcsname}
			{\noexpand\orcidicon}}
}

\date{\today}

\begin{document}

\preprint{UCI-HEP-TR-2023-04; FERMILAB-PUB-23-538-T-V}

\title{Constraints on fifth forces and ultralight dark matter \\ from OSIRIS-REx target asteroid Bennu}

\author{Yu-Dai Tsai\hspace{-1mm}\orcidA{}}
\email{yudait1@uci.edu}
\email{yt444@cornell.edu}
\affiliation{Department of Physics and Astronomy, University of California, 4129 Frederick Reines Hall, Irvine, CA 92697, USA \looseness=-1}
\affiliation{Fermi National Accelerator Laboratory (Fermilab), Batavia, IL 60510, USA}
\affiliation{Kavli Institute for Cosmological Physics (KICP), University of Chicago, Chicago, IL 60637, USA}

\author{Davide Farnocchia\hspace{-1mm}\orcidB{}}
\email{Davide.Farnocchia@jpl.nasa.gov}
\affiliation{Jet Propulsion Laboratory (JPL), California Institute of Technology, 4800 Oak Grove Drive, Pasadena, CA 91109, USA \looseness=-1}

\author{Marco Micheli\hspace{-1mm}\orcidC{}}
\email{marco.micheli@ext.esa.int}
\affiliation{European Space Agency (ESA) NEO Coordination Centre, Largo Galileo Galilei 1, 00044 Frascati (RM), Italy \looseness=-1}

\author{Sunny Vagnozzi\hspace{-1mm}\orcidD{}}
\email{sunny.vagnozzi@unitn.it}
\affiliation{Department of Physics, University of Trento, Via Sommarive 14, 38123 Povo (TN), Italy}
\affiliation{Trento Institute for Fundamental Physics and Applications (TIFPA)-INFN, Via Sommarive 14, 38123 Povo (TN), Italy \looseness=-1}

\author{Luca Visinelli\hspace{-1mm}\orcidE{}}
\email{luca.visinelli@sjtu.edu.cn}
\affiliation{Tsung-Dao Lee Institute (TDLI), 520 Shengrong Road, 201210 Shanghai, P.\ R.\ China}
\affiliation{School of Physics and Astronomy, Shanghai Jiao Tong University, 800 Dongchuan Road, 200240 Shanghai, P.\ R.\ China \looseness=-1}

\begin{abstract}
\noindent Using the OSIRIS-REx mission and ground-based tracking data for the asteroid Bennu, we derive new constraints on fifth forces and ultralight dark matter. The bounds we obtain are strongest for mediator masses $m \sim 10^{-18} - 10^{-17}\,{\rm eV}$, where we currently achieve the tightest bounds. Our limits can be translated to a wide class of models leading to Yukawa-type fifth forces, and we demonstrate how they apply to $U(1)_B$ dark photons and baryon-coupled scalars. Our results demonstrate the potential of asteroid tracking in probing well-motivated extensions of the Standard Model and ultralight dark matter satisfying the fuzzy dark matter constraints.
\end{abstract}

\maketitle

\section{Introduction}
\label{sec:introduction}

Anomalies in the trajectories of objects across the sky have often led to the discovery of new physical laws or celestial bodies. The planet Neptune was inferred based on the irregularities of the orbit of Uranus according to Newton’s theory, and General Relativity was first confirmed with its prediction of Mercury's anomalous precession~\cite{1859AnPar...5....1L,Einstein:1916vd}.
Besides planets, other bodies in the Solar Systems that are being tracked with increasing precision include near-Earth asteroids.
Currently, asteroids are tracked with plane-of-sky astrometric measures paired with line-of-sight studies by radar astrometry. Stellar occultations~\cite{2022A&A...658A..73F} and the Gaia satellite~\cite{2022arXiv220605561T} also provide highly-accurate astrometry. 

Asteroid tracking and planetary defense are crucial missions of the National Aeronautics and Space Administration (NASA) and the European Space Agency (ESA). Consequently, a wealth of data related to the orbits of these objects is available and can be used to probe fundamental physics, including new physics beyond the Standard Model of particle physics (SM). This is the path we take in the present work, focusing on data from the Origins, Spectral Interpretation, Resource Identification, Security, Regolith Explorer (OSIRIS-REx) mission. OSIRIS-REx is a NASA space mission designed to study the potentially hazardous asteroid (101955) Bennu. The spacecraft is equipped with several scientific instruments, including a camera suite, a laser altimeter, and a spectrometer, which allowed it to map Bennu's surface and study its composition, geology, and mineralogy, as well as improve the knowledge Bennu's future trajectory to reassess the probability of a future impact on Earth~\cite{2017SSRv..212..925L}. Launched in 2016, the OSIRIS-REx spacecraft arrived at the near-Earth asteroid Bennu and began conducting scientific observations and measurements in December 2018, before performing a Touch-and-Go (TAG) sample acquisition maneuver in October 2020, during which a sample of carbonaceous regolith of at least $60\,{\rm g}$ was collected and is expected to return to Earth for analysis in September 2023~\cite{2022Sci...377..285L}.
Overall, the OSIRIS-REx mission has provided an exciting opportunity for scientists to explore and study one of the most primitive objects in the solar system, and to gain new insights into the history and evolution of our cosmic neighborhood. The OSIRIS-REx tracking data are archived in the Small Bodies Node of the Planetary Data System, see Ref.~\cite{SBNPDS}. The data has been used to track to exquisite precision the trajectory of the asteroid Bennu, measure the Yarkovsky effect acting on it, and refine the long-term impact hazard for the asteroid~\cite{2021Icar..36914594F}.

Asteroid tracking can be invoked to test theories of gravity such as the validity of Newton's inverse square law. In fact, the laws of gravity would be modified if a new (ultra)light field exists in Nature, due to the capability for such a field to act as a ``fifth force'' and modify the orbits of larger bodies. The possibility of the existence of hidden fifth forces, in addition to those included in the SM, as well as the weakly coupled ultralight particles that mediate these forces, are topics of extreme importance in modern particle physics. Among others, these light particles are natural candidates for the DM and dark energy~\cite{Wilczek:1977pj,Peccei:1987mm,Wetterich:1987fm,Ratra:1987rm,Caldwell:1997ii,Carroll:1998zi,Zlatev:1998tr,Steinhardt:1999nw,Matos:1999et,Hu:2000ke,Wetterich:2002ic,Kim:2002tq,Khoury:2003rn,Marsh:2015xka,Hui:2016ltb,Ibe:2018ffn,Mocz:2019pyf,Ferreira:2020fam,Hui:2021tkt,Choi:2021aze}, and appear ubiquitously within string theory~\cite{Svrcek:2006yi,Arvanitaki:2009fg,Cicoli:2012sz,Visinelli:2018utg,Visinelli:2019qqu}. Fifth forces mediated by a new light bosonic field appear in extensions of the SM and, more generally, in beyond the SM (BSM) models that attempt to incorporate the observed DM and dark energy contents into a general framework. Many motivated BSM models in this sense have been studied in the literature, and include the gauged $U(1)_B$~\cite{Carone:1994aa,FileviezPerez:2010gw,Elor:2018twp}, $U(1)_{B-L}$~\cite{Davidson:1978pm,Mohapatra:1980qe,Davidson:1987mh,Escudero:2018fwn,Lin:2022xbu}, $L_{\mu}-L_{e, \tau}$~\cite{Foot:1990mn,He:1991qd,Escudero:2019gzq}, baryon-coupled scalar~\cite{Izaguirre:2014cza,Pospelov:2017kep,Blinov:2018vgc,Sibiryakov:2020eir,Paliathanasis:2023ttu}, and massive gravity~\cite{Hassan:2011hr,Hassan:2011zd,Bernus:2019rgl, Bernus:2020szc} models.

A wide range of probes have been used to search for and constrain hidden fifth forces and ultralight particles, including but not limited to laboratory and space tests~\cite{Fischbach:1985tk,DeRujula:1986ug,Talmadge:1988qz, Fischbach:1992fa,PhysRevD.50.3614,Hoyle:2004cw,Mota:2006fz,Brax:2007vm,Schlamminger:2007ht,Brax:2011hb,Wagner:2012ui,Burrage:2014oza,Foot:2014uba,Foot:2014osa,Brax:2015hma,Brax:2016did,TheMADMAXWorkingGroup:2016hpc,Perivolaropoulos:2016ucs,Burrage:2017qrf,Irastorza:2018dyq,Capolupo:2019xyd,Perivolaropoulos:2019vkb,Blanco:2019hah,Capolupo:2019peg,Braine:2019fqb,DiLuzio:2020wdo,KumarPoddar:2020kdz,Bloch:2020uzh,Vagnozzi:2021quy,Tsai:2021lly,Adams:2022pbo,Antypas:2022asj,Lambiase:2022ucu,Poddar:2023pfj,Elder:2023oar}, cosmological~\cite{Hlozek:2014lca,Baumann:2015rya,DEramo:2018vss,SimonsObservatory:2018koc,Poulin:2018dzj,Lambiase:2018lhs,Poulin:2018cxd,Vagnozzi:2019ezj,SimonsObservatory:2019qwx,Vagnozzi:2019kvw,Green:2019glg,EscuderoAbenza:2020cmq,Odintsov:2020iui,Rogers:2020ltq,Giare:2020vzo,Esteban:2021ozz,DiValentino:2021izs,Vagnozzi:2021gjh,Perivolaropoulos:2021jda,Xu:2021rwg,Ferlito:2022mok,Banerjee:2022era,DEramo:2022nvb,Bolton:2022hpt,Mazde:2022sdx,DiValentino:2022edq,Rogers:2023ezo,Oikonomou:2023kqi,Vanzan:2023gui} and astrophysical observations~\cite{Jain:2012tn,Giannotti:2015kwo,Foot:2016wvj,Caputo:2017zqh,Stott:2018opm,Roy:2019esk,Davoudiasl:2019nlo,Nojiri:2019riz,Nojiri:2019nar,Kumar:2020hgm,Nojiri:2020pqr,Creci:2020mfg,Khodadi:2020jij,Croon:2020oga,Stott:2020gjj,Desmond:2020nde,Caputo:2021efm,Khodadi:2021owg,Mehta:2021pwf,Caputo:2021eaa,Tsai:2021irw,Atamurotov:2021cgh,Calza:2021czr,Poddar:2021ose,Sakstein:2022tby,Dekens:2022gha,Vagnozzi:2022moj,Cannizzaro:2022xyw,Calza:2022ioe,Saha:2022hcd,Tsai:2022jnv,Jha:2022tdl,Unal:2023yxt,Oikonomou:2023bah,Parbin:2023zik,Calza:2023rjt}. Among others, the Lunar Laser Ranging (LLR) is monitoring the distance between the Moon and Earth with millimetric precision, placing strong bounds on the time variability of Newton's constant and on the strong equivalence principle~\cite{Williams:2004qba, Viswanathan:2017vob, Hofmann:2018myc}. Proof-of-principle studies have been discussed as possible ways to place model-dependent limits on a novel fifth force using asteroid and planetary precessions~\cite{Poddar:2020exe, Tsai:2021irw}. The data has also been used to obtain constraints on the local DM and cosmic neutrino densities~\cite{Tsai:2022jnv}. Planetary ephemerides released by the Observatory of Paris and the Co\^{t}e d’Azur Observatory (INPOP)~\cite{2017NSTIM.108.....V, 2019NSTIM.109.....F} have also been used to constrain fifth-force ranges~\cite{Fienga:2023ocw} and the effects of a finite graviton mass~\cite{Mariani:2023ubf}.

In this paper, we conduct the first-ever robust search for a hypothetical fifth force with precision asteroid tracking. To reach our goal, we utilize data for the orbit of Bennu from the OSIRIS-REx mission~\cite{2021Icar..36914594F}. We also considered a second asteroid, namely (99942) Apophis, which was extensively tracked with ground-based optical and radar telescopes from 2004 to 2021. Apophis will be visited by the extended OSIRIS-REx mission, OSIRIS-APEX~\cite{2022LPICo2681.2011D}, which may further improve the constraints reported here. We select a few simple and well-motivated realizations of the existing BSM models to demonstrate the power of our constraints. Nevertheless, it is straightforward to map the parametrization adopted here to other models that lead to a Yukawa-type fifth force among those mentioned above, e.g.,\ Refs.~\cite{Moffat:2005si, Clifton:2011jh}.

The rest of this paper is then organized as follows. In Sec.~\ref{sec:fifthforce}, we introduce the fifth force formulations and characterize the induced perturbations to the motions of asteroids. In Sec.~\ref{sec:analysis}, we briefly discuss our analysis method. Our results are presented and discussed in Sec.~\ref{sec:results}. Finally, in Sec.~\ref{sec:conclusions} we draw concluding remarks. Unless otherwise specified, throughout the paper, we use natural units ($\hbar = c = 1$), where $\hbar$ is the reduced Planck's constant and $c$ is the speed of light.

\section{Fifth forces and motion of celestial objects}
\label{sec:fifthforce}

A number of scenarios beyond the SM, including but not limited to those mentioned in Sec.~\ref{sec:introduction}, predict the existence of new (ultra)light particles, which would mediate a new, long-ranged, fifth force. For example, the SM contains a number of Abelian or $U(1)$ global symmetries, which are treated as ``accidental'' symmetries, which should be spoiled by the appearance of UV-completing operators such as those arising in quantum gravity~\cite{Giddings:1988cx, Coleman:1988tj}. For this, a $U(1)$ symmetry can be promoted to a gauge symmetry that is spontaneously broken below some UV energy scale, with the corresponding gauge boson $\phi$ in the case of a scalar field, or a vector field $A^{\prime}$ in the case of an axial vector gauge. Examples include models based on new (gauged) symmetries such as gauged baryon number conservation $U(1)_B$, $U(1)_{B-L}$, $L_{\mu}-L_{e, \tau}$, as well as baryon-coupled scalars.

A light scalar field $\phi$ of mass $m_\phi$ introduces a new channel through which particles can exchange momentum ${\bf q}$ and scatter with an amplitude $\vert \mathcal{M} \vert \propto 1/(q^2+m_\phi^2)$, where $q = |{\bf q}|$. The fifth force potential associated with this scattering amplitude is of the Yukawa type~\cite{Wagoner:1970vr, Talmadge:1988qz, Fischbach:1992fa, Will:2014kxa}, with the boson mass providing a cutoff over which the force extends, whereas the details concerning the force coupling depend on the charge and spin structure of the particles involved in the scattering process. Likewise, some models predict the fifth force to be mediated by a hidden vector particle $A^{\prime}$ of mass $m_{A^{\prime}}$, often referred to as a dark photon. A review of tests on the modification of gravity that includes Yukawa-type forces can be found in Ref.~\cite{Adelberger:2003zx}. When describing the methods, we will refer to a scalar field $\phi$ of mass $m_{\phi}$, although the results can also be extended to contemplate the vector case.

The presence of a long-range Yukawa-type force leads to deviations from Newton's inverse square law, which can be tested and constrained. In fact, given a specific theory that contains a gauge symmetry, any body would acquire a charge $Q$ associated with the breaking of the gauge symmetry, so that the mutual interaction between two bodies of charges $Q_1$ and $Q_2$ at a distance $r$ reads
\begin{equation}
    \label{eq:yukawa}
    V(r) = \mp (\hbar c)\frac{g^2}{4\pi}\frac{Q_1\,Q_2}{r} \exp \left (-\frac{m_\phi c}{\hbar}r \right ) \,,
\end{equation}
where $g$ is the coupling strength associated with the gauge field. Given the position ${\bf r}$ of the celestial object with respect to the Sun, the central potential only depends on the radial coordinate $r = |{\bf r}|$. This is a very generic parametrization that encompasses a number of interesting scenarios. This result has profound implications for the motion of celestial objects around the Sun.

We then specialize the discussion to an asteroid of mass $M_{\ast}$ orbiting in the solar system under the influence of the Sun of mass $M_{\odot}$. For instance, in the case of a gauged $U(1)_B$ model, the charges under the specific gauge group of the Sun and the celestial body result in $Q_{\odot}=M_{\odot}/m_p$ and $Q_{\ast}=M_{\ast}/m_p$ respectively, with $m_p$ denoting the proton mass. The conversion between $g$ and $\widetilde{\alpha}$ can be found in Refs.~\cite{Schlamminger:2007ht,Wagner:2012ui}. The effect of the fifth force in Eq.~\eqref{eq:yukawa} due to a mediator of mass $m_\phi$ through the central potential in the model proposed translates as
\begin{eqnarray}
    V(r) = \widetilde{\alpha}\,\frac{GM_\odot M_\ast}{r}\exp \left ( -\frac{r}{\lambda} \right ) \,,
    \label{eq:vryukawa}
\end{eqnarray}
where $\widetilde{\alpha}$ characterizes the strength of the fifth force, and $\lambda=\hbar/m_\phi c$ is the fifth force range. 

In this work, we shall remain as agnostic as possible for what concerns the microscopical origin of the fifth force, and will simply parameterize it in terms of $\widetilde{\alpha}$ and $\lambda$ as appearing in Eq.~\eqref{eq:vryukawa}. However, we emphasize that it is always possible to translate these constraints to the corresponding limits in terms of the mass and coupling of an ultralight particle, once a specific underlying fifth force model is chosen. Note that in the limit $m_\phi \to 0$, and therefore $\lambda \to +\infty$, the effects of the potential in Eq.~\eqref{eq:vryukawa} translate into a modification of Newton's constant, $G \to G(1+\widetilde\alpha)$. The symmetries of the problem (and in particular, the potential being central) indicate that the motion is typically planar to a very good approximation, so the coordinate system is fixed such that the polar angle is $\theta=\pi/2$.

It is useful to consider the reciprocal coordinate $u=1/r$, in terms of which the equation of motion for the celestial object, including General Relativity (GR) corrections and temporarily restoring SI units, is given by~\cite{Poddar:2020exe,Tsai:2021irw}
\begin{equation}
    \label{eq:equationsofmotion}
    \frac{{\rm d}^2 u}{{\rm d}\varphi^2} + u = \frac{GM_\odot}{L^2}+\frac{3GM_\odot}{c^2} u^2 + \widetilde{\alpha}\frac{G M_\odot}{L^2} \left ( 1+\frac{1}{\lambda u} \right) e^{-\frac{1}{\lambda u}}\,,
\end{equation}
where $\varphi$ is the azimuthal angle that parameterizes the motion of the celestial object, $L$ denotes the orbital angular momentum per unit mass, whereas the first, second, and third set of terms on the right-hand side describe the effects of Newtonian physics, GR corrections, and the fifth force, respectively.

In Newtonian physics, the motion of a bound object in an isolated two-body system traces out a fixed elliptical orbit; in particular, the semimajor axis of the ellipse is fixed in space. However, the last two sets of terms on the right-hand side of Eq.~(\ref{eq:equationsofmotion}) lead to a precession (rotation) of the celestial object's perihelion as it revolves around the Sun. In other words, $u$ changes as the azimuthal angle is shifted by $\varphi \to \varphi+2\pi n$, or equivalently, the celestial object will return to its perihelion after one precession at an azimuthal angle which differs from $2\pi$ by a quantity $\Delta \varphi$. This precession leads to potentially observable effects in the motion of celestial objects, including but not limited to asteroids and planets.

The Newtonian solution to Eq.~(\ref{eq:equationsofmotion}) for an asteroid in a closed orbit of eccentricity $\mathsf{e}$ is $u_0(\varphi)=M_{\odot}(1+\mathsf{e}\cos\varphi)/L^2$. To determine the perihelion precession (measured from a fixed reference direction) per orbital period as a function of the fifth force strength and range, we can numerically solve Eq.~(\ref{eq:equationsofmotion}) for $u(\varphi)$, expanding perturbatively around the Newtonian solution $u_0(\varphi)$, and deriving the shift in the $\varphi$-period relative to $2\pi$, $\Delta\varphi$, from this solution (see Ref.~\cite{Tsai:2021irw} and Appendix~B of Ref.~\cite{Tsai:2022jnv} for more details). Dividing the shift
by the orbital period, one obtains the perihelion precession rate $\Delta\dot{\varphi}$.

While a closed-form expression for the contribution to the perihelion precession from fifth forces described by the potential in Eq.~(\ref{eq:vryukawa}) is generally not available, an approximate expression can be obtained in the limit of a very light mediator $m_\phi \ll \hbar/ac$, i.e., \ when the semimajor axis $a$ is much smaller than the fifth force range. In this case, one finds (see, e.g., \ Ref.~\cite{Tsai:2021irw}):
\begin{eqnarray}
\vert \Delta\dot{\varphi} \vert \simeq \frac{2\pi\widetilde{\alpha}}{1+\widetilde{\alpha}} \left ( \frac{am_\phi c}{\hbar} \right ) ^2 \left ( 1-\mathsf{e} \right ) = \frac{2\pi\widetilde{\alpha}}{1+\widetilde{\alpha}} \left ( \frac{a}{\lambda} \right ) ^2 \left ( 1-\mathsf{e} \right ),\,
\label{eq:precessionlowmass}
\end{eqnarray}
which correctly vanishes in the limit $m_\phi \to 0$, reflecting the fact that this limit simply recovers the inverse square law (or equivalently a $\propto 1/r$ potential, albeit with a different value of the gravitational constant), which does not lead to precession. Note that the perihelion precession $\vert \Delta\dot{\varphi} \vert$ increases with the square of the semimajor axis under this limit, which indicates the benefit of studying objects at a relatively large distance from the Sun such as Trans-Neptunian objects (TNOs). Such a dataset could also improve the determination of the DM density within the solar system performed in Ref.~\cite{Tsai:2022jnv}, for which the perihelion precession benefits from an additional power of $a$ compared to our case.

We utilize the state-of-the-art asteroidal orbital determination, for which meter-level tracking data can be available, to provide new constraints on fifth forces.
To do so, we employ the Comet and Asteroid Orbit Determination Package developed and maintained by NASA Jet Propulsion Lab (JPL), whose force models account for N-body Newtonian and relativistic gravity, oblateness terms, and nongravitational perturbations~\cite{2021Icar..36914594F}, supplemented with the fifth force we intend to constrain. These aspects will be discussed in more detail in Sec.~\ref{sec:analysis}.

\section{Data and Analysis}
\label{sec:analysis}

The trajectory of the asteroid Bennu has been extremely well constrained by a combination of ground-based optical and radar astrometric data collected ever since its discovery in 1999. The level to which Bennu's ephemeris is constrained was significantly improved by X-band radiometric and optical navigation tracking data collected during asteroid proximity operations by the OSIRIS-REx mission, from its arrival in December 2018 to sample collection in October 2020.

In this work, we make use of the same dataset used in the earlier Bennu ephemeris and hazard assessment analysis of Ref.~\cite{2021Icar..36914594F}. In particular, this includes:
\begin{itemize}
\item ground-based optical astrometry data, including 489 right ascension and declination observations from 1999-09-11 to 2018-05-15~\footnote{\url{https://www.minorplanetcenter.org/db_search/show_object?utf8=?&object_id=Bennu}};
\item all available ground-based radar astrometry data, collected during three close encounters in September 1999, September 2005, and September 2011 by the Arecibo and Goldstone radar stations, and comprising 7 Doppler and 22 delay measurements~\footnote{\url{https://ssd.jpl.nasa.gov/?grp=num&fmt=html&radar=}};
\item 36 geocentric pseudo-range points for Bennu's barycenter between 2019-01-03 16:56:56 UTC and 2020-10-03 19:14:18 UTC, derived from the OSIRIS-REx high-gain antenna tracking data in Ref.~\cite{2021Icar..36914594F} (see Tab.~1 therein), and for which a $15\,{\rm ns}$ uncertainty is assumed and corresponds to an uncertainty of $2\,{\rm m}$ for the radial distance between Bennu and Earth.~\footnote{These points have been obtained using tracking data obtained during the mission orbital phases Orbital A, B, C, and R, Reconnaissance B and C, Rehearsal, and pre-TAG (during which the OSIRIS-REx spacecraft was in a closed orbit around Bennu), and selecting independent, maneuver-free arcs of approximately 10 days, see Ref.~\cite{2021Icar..36914594F} for more details.}
\end{itemize}
All the above data was used to compute an orbit solution, as in Ref.~\cite{2021Icar..36914594F}.

The pseudo-range points discussed above lead to a significant improvement in the determination of Bennu's trajectory. This, in turn, imposes stringent requirements on the fidelity of the underlying force model used to fit the trajectory. We adopt the same unprecedented, high-fidelity model developed for the purposes of the earlier Bennu ephemeris and hazard assessment analysis of Ref.~\cite{2021Icar..36914594F}. This model accounts for (relativistic) gravitational effects from the Sun, the eight planets, Pluto, and the Moon (modeled through a first-order parametrized post-Newtonian N-body formulation, also known as Einstein–Infeld–Hoffman formulation, see Refs.~\cite{Einstein:1938yz,Will:1993hxu}), point-mass Newtonian gravitational effects from 343 small-body perturbers, and gravitational effects from the Earth's oblateness, as well as a number of non-gravitational perturbation effects which include the Yarkovsky effect, solar radiation pressure, and Poynting-Robertson drag. The orbit solution is determined using the JPL Comet and Asteroid Orbit Determination Package, and employing the DIVA variable order Adams integrator to quadruple precision, with integration tolerance of $10^{-18}$. For further details on the underlying force model and integration precision requirements, we refer the reader to Ref.~\cite{2021Icar..36914594F}.

To derive fifth force constraints, the force model of Ref.~\cite{2021Icar..36914594F} is further expanded to include the acceleration associated with the fifth force, obtained by differentiating Eq.~\eqref{eq:vryukawa} and dividing by the asteroid mass $M_\ast$:
\begin{eqnarray}
{\bf a}({\bf r}) = \widetilde{\alpha}\frac{GM_\odot}{r^3} e^{-\frac{r}{\lambda}} \left ( 1 + \frac{r}{\lambda} \right ) {\bf r}\,.
\label{eq:fifthforce}
\end{eqnarray}
Note that in our force model, we apply the above acceleration term only to Bennu, not the other bodies: a posteriori, given the very stringent upper limits on the fifth force strength, we expect this to be a valid approximation for the range of parameter space explored.

As in Ref.~\cite{2021Icar..36914594F}, the model parameters varied in the fit to determine the orbit solution are Bennu's heliocentric orbital elements (eccentricity, perihelion distance, time of perihelion TDB, longitude of node, argument of perihelion, and inclination) at osculating epoch 2011 January 1.0 TDB, Bennu's bulk density, Bennu's area-to-mass ratio, the masses of the 343 small-body perturbers, and a constant delay bias for the pseudo-range points. To these parameters, we add the parameters characterizing the fifth force, i.e., its range $\lambda$ and strength $\widetilde{\alpha}$.

We perform a least-squares fit to Bennu optical and radar astrometry, and pseudo-range points discussed previously, explicitly varying all the parameters mentioned previously, except for $\lambda$, which we fix to 51 logarithmically spaced values between $10^{-2}\,{\rm au}$ and $10^3\,{\rm au}$. For each of these fixed values of $\lambda$, we estimate all the model parameters, and focus on the marginal distribution for $\widetilde{\alpha}$, which ultimately is the parameter we care about.  As discussed below, we find that the marginal distribution for $\widetilde{\alpha}$ is statistically compatible with $\widetilde{\alpha}=0$, and by extension, that the marginal distributions for all the other parameters are in excellent agreement with their earlier estimates in Ref.~\cite{2021Icar..36914594F}, whose force model assumed no fifth force. In other words, there is no evidence in the data for the presence of a fifth force affecting the motion of Bennu. For this reason, in the following (and in particular in Fig.~\ref{fig:apophis_bennu_constraints_vector_scalar}), we adopt the common choice of only reporting $2\sigma$ upper limits on $\widetilde{\alpha}$, determined from its marginal distribution.

\section{Results and Discussions}
\label{sec:results}

\begin{figure*}[!ht]
\centering
\includegraphics[width=0.9\textwidth]{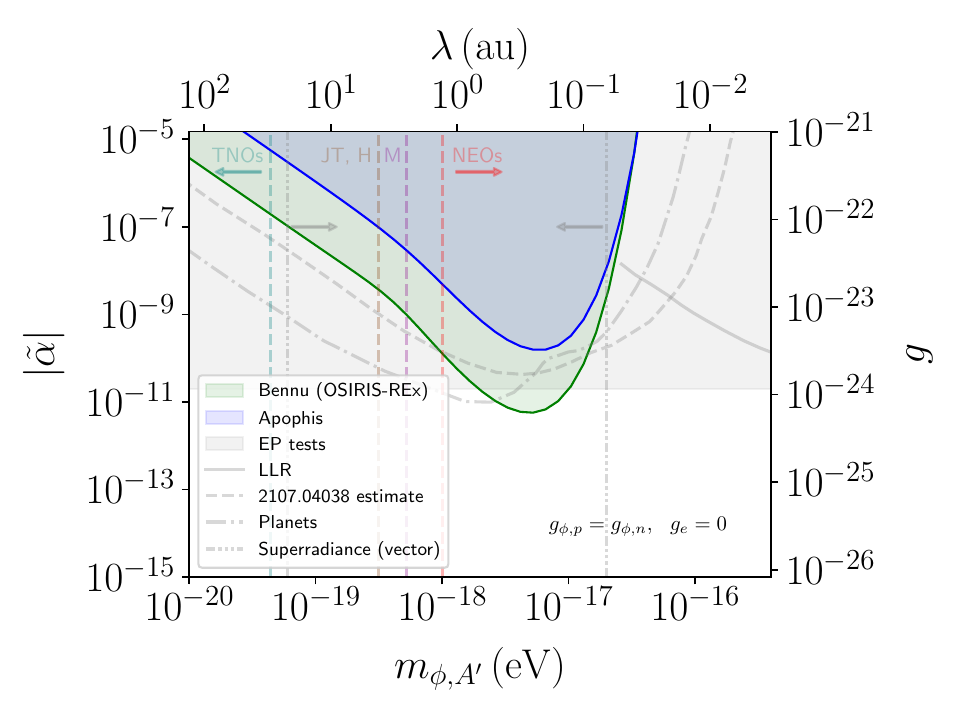} \qquad
\caption{Upper limits on the coupling strength of $U(1)_B$ dark photons $A^{\prime}$ and baryon-coupled scalars $\phi$ with $g_{\phi,p}=g_{\phi,n}$ and $g_e=0$, as a function of the particle mass or, equivalently, its Compton wavelength. The limits have been obtained by analyzing the orbits of Bennu (green) and Apophis (blue), and the shaded regions above the respective curves are excluded. Also shown are upper limits from other probes: equivalence principle (EP) tests (grey shaded area is excluded)~\cite{Touboul:2017grn,Fayet:2018cjy,MICROSCOPE:2022doy}, lunar laser ranging (LLR, grey solid curve)~\cite{Williams:2004qba, Viswanathan:2017vob, Hofmann:2018myc}, planetary precession (grey dash-dotted curve)~\cite{Poddar:2020exe}, vector superradiance (excluding the region enclosed between the two grey dash-dotdotdotted lines)~\cite{Brito:2015oca,Baryakhtar:2017ngi,Cardoso:2017kgn}, and the earlier sensitivity reach estimated from the precessions of 9 asteroids (grey dashed curve)~\cite{Tsai:2021irw}. The four colored dashed vertical lines indicate the mass ranges indicatively probed by other classes of objects as per the colored labels: Jupiter Trojans (JT) and Hildas (H) in brown, main-belt asteroids (M) in magenta, Near-Earth Objects (NEOs) in red, and Trans-Neptunian Objects (TNOs) in teal.}
\label{fig:apophis_bennu_constraints_vector_scalar}
\end{figure*}

The outcome of the analysis detailed above is shown in Fig.~\ref{fig:apophis_bennu_constraints_vector_scalar}, where we consider a fifth force model in which the mediator is either a dark photon $A^{\prime}$ of mass $m_{A^{\prime}}$ or a baryon-coupled scalar field $\phi$ of mass $m_\phi$, arising from a $U(1)_B$ local gauged symmetry. In this model, only the baryons feel the presence of the fifth force so that the couplings of the mediator to the proton and neutron are equal, $g_{p}=g_{n}$, while the coupling to the electron is zero, $g_e=0$. We report $2\sigma$ upper limits on the dimensionless coupling $\vert \widetilde\alpha \vert$ versus the mediator mass $m_{\phi,A^{\prime}}$ (lower horizontal axis), or equivalently the range of the fifth force $\lambda$ (upper horizontal axis), using data from the OSIRIS-REx mission tracking the asteroid Bennu (green solid curve, with the green shaded region above the curve excluded). To provide a comparison, we ran the same analysis on the asteroid Apophis, using data available from optical and ground-based radar telescopes tracking its trajectory~\cite{2022LPICo2681.2007F} (blue solid curve, once more with the blue shaded region above the curve excluded). Given the significantly enhanced sensitivity of the OSIRIS-REx mission at the meter-level scale, the constraints obtained when fitting this dataset (for Bennu's orbit) are currently much stronger than the constraints obtained with Apophis for $\lambda \gtrsim 3 \times 10^{-2}\,{\rm au}$, which is precisely the region we are featuring in Fig.~\ref{fig:apophis_bennu_constraints_vector_scalar}.

The sensitivity we find is particularly strong in the region $10^{-18}\,{\rm eV} \lesssim m_{\phi,A^{\prime}} \lesssim 10^{-16}\,{\rm eV}$, and is strongest for $m_{\phi,A^{\prime}} \sim {\cal O}(10^{-17})\,{\rm eV}$, which corresponds to a length scale $\lambda \sim {\cal O}(0.1)\,{\rm au}$. This is due to a combination of Bennu's eccentricity $\mathsf{e} \approx 0.20375$ and its semimajor axis $a=1.1264\,{\rm au}$~\cite{2021Icar..36914594F}, which allows us to set leading constraints in the range $10^{-18}\,{\rm eV} \lesssim m_{\phi,A^{\prime}} \lesssim 10^{-17}\,{\rm eV}$ when compared to planetary probes. In fact, the tracking of an asteroid orbiting with semimajor axis $a$ and no eccentricity would translate into a fifth force constrain that peaks at the mass $m_{\phi,A^{\prime}} \sim \hbar c/a$, which would correspond to $\sim 10^{-18}\,$eV for an asteroid of semimajor axis same as Bennu's. The inclusion of the eccentricity makes the asteroid probe regions of the orbit that are smaller than $a$, with the closest approach to the Sun being $r_{\rm min} = a(1-\mathsf{e})/(1+\mathsf{e})$: this leads to the constraint shifting towards higher values of the mediator mass. The same reasoning applies to the results obtained with Apophis tracking data, whose eccentricity and semimajor axis are $\mathsf{e} \approx 0.1912$ and $a=0.9224\,{\rm au}$ respectively~\cite{2018Icar..300..115B}.

For length scales larger than $\hbar/m_{\phi,A^{\prime}}c$, the Yukawa potential altering Eq.~\eqref{eq:equationsofmotion} can be expanded to yield a correction $\propto \lambda/\sqrt{\alpha}$. A fit to the results obtained gives
\begin{equation}
    10^{-13}\,\frac{\lambda/{\rm km}}{\sqrt{\alpha}} > 1.02\,,
\end{equation}
where the result is valid at $2\sigma$ and for length scales $\lambda \gg\,$au. Similar results have been obtained using INPOP ephemerides leading to $10^{-13}\,(\lambda/{\rm km})/\sqrt{\alpha} > 1.83$~\cite{Bernus:2019rgl} and $10^{-13}\,(\lambda/{\rm km})/\sqrt{\alpha} > 3.93$~\cite{Bernus:2020szc} at $1\sigma$ and for a positive coupling strength $\alpha$. For a comparison, the bound we obtained from using the Apophis ephemeris is $10^{-13}\,(\lambda/{\rm km})/\sqrt{\alpha} > 4.79$ at $2\sigma$.

Also shown in Fig.~\ref{fig:apophis_bennu_constraints_vector_scalar} are fifth force sensitivities obtained from other probes, including planetary motion (grey dash-dotted curve), which peaks for $m_{\phi,A^{\prime}} \sim {\cal O}(10^{-18})\,{\rm eV}$, and was obtained by searching for an anomalous precession in the orbits of solar system planets~\cite{KumarPoddar:2020kdz} (see also Refs.~\cite{Iorio:2007gq,DeMartino:2018yqf,Sun:2019ico,Davis:2019ltc,Benisty:2022txp}). The grey solid curve labeled ``LLR'' shows the sensitivity obtained from LLR probes of the anomalous precession of the Moon~\cite{Williams:2004qba, Viswanathan:2017vob, Hofmann:2018myc}.~\footnote{Other measurements based on the orbits of the Moon and the LAGEOS satellite, see e.g.\ Ref.~\cite{DeRujula:1986ug}, place constraints on much shorter length scales, which are not shown in Fig.~\ref{fig:apophis_bennu_constraints_vector_scalar}.} The grey shaded area is excluded by the MICROSCOPE mission~\cite{Touboul:2017grn,Fayet:2018cjy,MICROSCOPE:2022doy}, which provides an accurate test of the weak equivalence principle (EP), improving over laboratory torsion balance experiments and atom interferometry measurements. Finally, the existence of a vector mediator in the mass range $6 \times 10^{-20}\,{\rm eV} \lesssim m_{A^{\prime}} \lesssim 2 \times 10^{-17}\,{\rm eV}$ has been constrained by considerations on black hole superradiance~\cite{Baryakhtar:2017ngi} using spin measurements of supermassive black holes (SMBHs) from X-ray reflection spectroscopy~\cite{Reynolds:2013qqa}. We stress that these limits (enclosed between the two grey dash-dot-dot-dotted lines, as indicated by the two grey arrows, and only applying to the dark photon) should be interpreted with some caution for a number of reasons, including uncertainties inherent to the formation and spin evolution of SMBHs, the scarcity/reliability of other spin and mass measurements~\cite{Denney:2008gk,Shen:2013pea,Vagnozzi:2020quf}, as well as other considerations pertaining to the evolution of the superradiant instability in the presence of competing environmental effects (see e.g.\ Refs.~\cite{Arvanitaki:2014wva,Brito:2014wla,Roy:2021uye,Hui:2022sri,Chen:2022nbb}).

For reference, we also compare our results to those obtained from the earlier qualitative sensitivity analysis for a sample of nine other asteroids in Ref.~\cite{Tsai:2021irw}, estimated solely from the precision in measuring the semimajor axes, eccentricities, and perihelia precessions of the asteroids whose properties were studied in Ref.~\cite{Verma:2017ywb} (grey dashed curve). The bounds obtained in the present work are significantly more robust than those obtained in this earlier qualitative analysis, given that the current analysis results from a direct fit to high-quality orbital tracking and constraints on the distance between Bennu and Earth, rather than condensing all of the information on the asteroid orbits into a single number, namely its orbital precession. Moreover, the software and force model used here (based on Ref.~\cite{2021Icar..36914594F}) accounts for various non-gravitational effects, in addition to gravitational perturbations exerted by planets and the most massive asteroids, as a result significantly increasing the fidelity of the final constraints we have reported.

Finally, it is worth noting that modeling assumptions can potentially affect the estimates we have presented. When analyzing DM constraints from OSIRIS-REx data \cite{Tsai:2022jnv}, the most significant effect in this sense was found to be caused by the choice of planetary ephemeris version, and we find that the same is true here. In particular, when switching from the DE424~\cite{SSD} to the DE440~\cite{2021AJ....161..105P} ephemeris, we observe $0.1$ to $1.9\sigma$ shifts in the best-fit values of $\widetilde{\alpha}$ (depending on the underlying value of $\lambda$). To capture this uncertainty floor, in Fig.~\ref{fig:apophis_bennu_constraints_vector_scalar}, we have conservatively decided to report the formal $2\sigma$ upper limits, which therefore safely encompass the ephemeris-associated uncertainty budget.

\section{Summary}
\label{sec:conclusions}

We set new, leading constraints on hidden fifth forces and ultralight dark matter using OSIRIS-REx tracking data for the asteroid Bennu. Based on its size and probability of future impacts with Earth, Bennu is one of the most potentially hazardous of all the currently known near-Earth asteroids, which is one of the reasons why meter-level tracking data is available. While we considered a model for the ultralight mediators for concreteness, our constraints can be translated into a wide class of models featuring Yukawa-type interaction. Our robust constraints on the fifth force strength are stronger than the existing laboratory and space tests bounds for the mediator mass range $m_{\phi,A^{\prime}} \sim 10^{-18} - 10^{-17}\,{\rm eV}$. This result extends to hypothetical light mediators to include models in which the dark matter is extremely light or ``fuzzy,'' which would also be subject to these bounds obtained through asteroid tracking.

The results presented here will be improved further in future work thanks to i) the ephemeris that will be reported by the future OSIRIS-APEX mission in tracking Apophis~\cite{2022LPICo2681.2011D}; ii) the inclusion of the data from tracking all other NEOs, TNOs, Trojans, Hildas, and main-belt asteroids. Including different classes of asteroids and solar-system objects enables the coverage of more orbital configurations, providing more comprehensive constraints on the fifth-force mediator masses that roughly correspond to the inverse of their semi-major axes. 
Further improvements on the data side may be achieved through the employment of quantum technologies, including technologies similar to the Deep Space Atomic Clocks~\cite{DSAC2021}, for example, which have also been considered to study dark matter (DM) and gravitational waves~\cite{Tsai:2021lly,Fedderke:2021kuy,Tsai:2022jnv}, among other science targets.

\section*{Acknowledgements}
\noindent We thank Slava G.\ Turyshev, Cedric Delaunay, Olivier Minazzoli, Yotam Soreq, and Youjia Wu for useful discussions. Y.D.T.\ is supported by the U.S.\ National Science Foundation (NSF) Theoretical Physics Program, Grant No.~PHY-1915005. Y.D.T.\ also thanks the Institute for Nuclear Theory at the University of Washington for its kind hospitality and stimulating research environment. This research was partly supported by the INT's U.S.\ Department of Energy grant No.~DE-FG02-00ER41132. This work was partially performed at the Aspen Center for Physics, supported by National Science Foundation grant No.~PHY-2210452. The National Science Foundation under Grant No.~NSF PHY-1748958 partly supported this research. D.F.\ conducted this research at the Jet Propulsion Laboratory, California Institute of Technology, under a contract with the National Aeronautics and Space Administration (80NM0018D0004). S.V.\ is partially supported by the Istituto Nazionale di Fisica Nucleare (INFN) through the CSN4 Iniziativa Specifica FieLds And Gravity (FLAG). L.V.\ acknowledges the Galileo Galilei Institute for Theoretical Physics in Florence (Italy), the INFN Frascati National Laboratories near Rome (Italy), and the Weinberg Institute for Theoretical Physics at the University of Texas in Austin, TX (USA) for hospitality during the completion of this work. This publication is based upon work from COST Actions CA21106 -- ``COSMIC WISPers in the Dark Universe: Theory, astrophysics and experiments (COSMIC WISPers)'' and CA21136 --``Addressing observational tensions in cosmology with systematics and fundamental physics (CosmoVerse)'', both supported by COST (European Cooperation in Science and Technology).

\bibliographystyle{JHEP}
\bibliography{asteroidosirixrex}

\providecommand{\href}[2]{#2}\begingroup\raggedright\begin{thebibliography}{100}

\bibitem{1859AnPar...5....1L}
U.J.~{Le Verrier}, \emph{{Theorie du mouvement de Mercure}}, {\emph{Annales de
  l'Observatoire de Paris} {\bfseries 5} (1859) 1}.

\bibitem{Einstein:1916vd}
A.~Einstein, \emph{{The Foundation of the General Theory of Relativity}},
  \href{https://doi.org/10.1002/andp.200590044}{\emph{Annalen Phys.} {\bfseries
  49} (1916) 769}.

\bibitem{2022A&A...658A..73F}
J.F.~{Ferreira}, P.~{Tanga}, F.~{Spoto}, P.~{Machado} and D.~{Herald},
  \emph{{Asteroid astrometry by stellar occultations: Accuracy of the existing
  sample from orbital fitting}},
  \href{https://doi.org/10.1051/0004-6361/202141753}{\emph{Astron. Astrophys.}
  {\bfseries 658} (2022) A73}
  [\href{https://arxiv.org/abs/2201.03892}{{\ttfamily 2201.03892}}].

\bibitem{2022arXiv220605561T}
P.~{Tanga}, T.~{Pauwels}, F.~{Mignard}, K.~{Muinonen}, A.~{Cellino}, P.~{David}
  et~al., \emph{{Data Release 3: the Solar System survey}},
  \href{https://doi.org/10.48550/arXiv.2206.05561}{\emph{arXiv e-prints} (2022)
  arXiv:2206.05561} [\href{https://arxiv.org/abs/2206.05561}{{\ttfamily
  2206.05561}}].

\bibitem{2017SSRv..212..925L}
D.S.~{Lauretta}, S.S.~{Balram-Knutson}, E.~{Beshore}, W.V.~{Boynton},
  C.~{Drouet d'Aubigny}, D.N.~{DellaGiustina} et~al., \emph{{OSIRIS-REx: Sample
  Return from Asteroid (101955) Bennu}},
  \href{https://doi.org/10.1007/s11214-017-0405-1}{\emph{Space Science Reviews}
  {\bfseries 212} (2017) 925}
  [\href{https://arxiv.org/abs/1702.06981}{{\ttfamily 1702.06981}}].

\bibitem{2022Sci...377..285L}
D.S.~{Lauretta}, C.D.~{Adam}, A.J.~{Allen}, R.L.~{Ballouz}, O.S.~{Barnouin},
  K.J.~{Becker} et~al., \emph{{Spacecraft sample collection and subsurface
  excavation of asteroid (101955) Bennu}},
  \href{https://doi.org/10.1126/science.abm1018}{\emph{Science} {\bfseries 377}
  (2022) 285}.

\bibitem{SBNPDS}
N.J.P.L.~(JPL), ``Small bodies node of the planetary data system.''
  \url{https://sbn.psi.edu/pds/resource/orex/}, 2022.

\bibitem{2021Icar..36914594F}
D.~{Farnocchia}, S.R.~{Chesley}, Y.~{Takahashi}, B.~{Rozitis},
  D.~{Vokrouhlick{\'y}}, B.P.~{Rush} et~al., \emph{{Ephemeris and hazard
  assessment for near-Earth asteroid (101955) Bennu based on OSIRIS-REx data}},
  \href{https://doi.org/10.1016/j.icarus.2021.114594}{\emph{Icarus} {\bfseries
  369} (2021) 114594}.

\bibitem{Wilczek:1977pj}
F.~Wilczek, \emph{{Problem of Strong $P$ and $T$ Invariance in the Presence of
  Instantons}}, \href{https://doi.org/10.1103/PhysRevLett.40.279}{\emph{Phys.
  Rev. Lett.} {\bfseries 40} (1978) 279}.

\bibitem{Peccei:1987mm}
R.D.~Peccei, J.~Sola and C.~Wetterich, \emph{{Adjusting the Cosmological
  Constant Dynamically: Cosmons and a New Force Weaker Than Gravity}},
  \href{https://doi.org/10.1016/0370-2693(87)91191-9}{\emph{Phys. Lett. B}
  {\bfseries 195} (1987) 183}.

\bibitem{Wetterich:1987fm}
C.~Wetterich, \emph{{Cosmology and the Fate of Dilatation Symmetry}},
  \href{https://doi.org/10.1016/0550-3213(88)90193-9}{\emph{Nucl. Phys. B}
  {\bfseries 302} (1988) 668}
  [\href{https://arxiv.org/abs/1711.03844}{{\ttfamily 1711.03844}}].

\bibitem{Ratra:1987rm}
B.~Ratra and P.J.E.~Peebles, \emph{{Cosmological Consequences of a Rolling
  Homogeneous Scalar Field}},
  \href{https://doi.org/10.1103/PhysRevD.37.3406}{\emph{Phys. Rev. D}
  {\bfseries 37} (1988) 3406}.

\bibitem{Caldwell:1997ii}
R.R.~Caldwell, R.~Dave and P.J.~Steinhardt, \emph{{Cosmological imprint of an
  energy component with general equation of state}},
  \href{https://doi.org/10.1103/PhysRevLett.80.1582}{\emph{Phys. Rev. Lett.}
  {\bfseries 80} (1998) 1582}
  [\href{https://arxiv.org/abs/astro-ph/9708069}{{\ttfamily
  astro-ph/9708069}}].

\bibitem{Carroll:1998zi}
S.M.~Carroll, \emph{{Quintessence and the rest of the world}},
  \href{https://doi.org/10.1103/PhysRevLett.81.3067}{\emph{Phys. Rev. Lett.}
  {\bfseries 81} (1998) 3067}
  [\href{https://arxiv.org/abs/astro-ph/9806099}{{\ttfamily
  astro-ph/9806099}}].

\bibitem{Zlatev:1998tr}
I.~Zlatev, L.-M.~Wang and P.J.~Steinhardt, \emph{{Quintessence, cosmic
  coincidence, and the cosmological constant}},
  \href{https://doi.org/10.1103/PhysRevLett.82.896}{\emph{Phys. Rev. Lett.}
  {\bfseries 82} (1999) 896}
  [\href{https://arxiv.org/abs/astro-ph/9807002}{{\ttfamily
  astro-ph/9807002}}].

\bibitem{Steinhardt:1999nw}
P.J.~Steinhardt, L.-M.~Wang and I.~Zlatev, \emph{{Cosmological tracking
  solutions}}, \href{https://doi.org/10.1103/PhysRevD.59.123504}{\emph{Phys.
  Rev. D} {\bfseries 59} (1999) 123504}
  [\href{https://arxiv.org/abs/astro-ph/9812313}{{\ttfamily
  astro-ph/9812313}}].

\bibitem{Matos:1999et}
T.~Matos, F.S.~Guzman and L.A.~Urena-Lopez, \emph{{Scalar field as dark matter
  in the universe}},
  \href{https://doi.org/10.1088/0264-9381/17/7/309}{\emph{Class. Quant. Grav.}
  {\bfseries 17} (2000) 1707}
  [\href{https://arxiv.org/abs/astro-ph/9908152}{{\ttfamily
  astro-ph/9908152}}].

\bibitem{Hu:2000ke}
W.~Hu, R.~Barkana and A.~Gruzinov, \emph{{Cold and fuzzy dark matter}},
  \href{https://doi.org/10.1103/PhysRevLett.85.1158}{\emph{Phys. Rev. Lett.}
  {\bfseries 85} (2000) 1158}
  [\href{https://arxiv.org/abs/astro-ph/0003365}{{\ttfamily
  astro-ph/0003365}}].

\bibitem{Wetterich:2002ic}
C.~Wetterich, \emph{{Probing quintessence with time variation of couplings}},
  \href{https://doi.org/10.1088/1475-7516/2003/10/002}{\emph{JCAP} {\bfseries
  10} (2003) 002} [\href{https://arxiv.org/abs/hep-ph/0203266}{{\ttfamily
  hep-ph/0203266}}].

\bibitem{Kim:2002tq}
J.E.~Kim and H.P.~Nilles, \emph{{A Quintessential axion}},
  \href{https://doi.org/10.1016/S0370-2693(02)03148-9}{\emph{Phys. Lett. B}
  {\bfseries 553} (2003) 1}
  [\href{https://arxiv.org/abs/hep-ph/0210402}{{\ttfamily hep-ph/0210402}}].

\bibitem{Khoury:2003rn}
J.~Khoury and A.~Weltman, \emph{{Chameleon cosmology}},
  \href{https://doi.org/10.1103/PhysRevD.69.044026}{\emph{Phys. Rev. D}
  {\bfseries 69} (2004) 044026}
  [\href{https://arxiv.org/abs/astro-ph/0309411}{{\ttfamily
  astro-ph/0309411}}].

\bibitem{Marsh:2015xka}
D.J.E.~Marsh, \emph{{Axion Cosmology}},
  \href{https://doi.org/10.1016/j.physrep.2016.06.005}{\emph{Phys. Rept.}
  {\bfseries 643} (2016) 1} [\href{https://arxiv.org/abs/1510.07633}{{\ttfamily
  1510.07633}}].

\bibitem{Hui:2016ltb}
L.~Hui, J.P.~Ostriker, S.~Tremaine and E.~Witten, \emph{{Ultralight scalars as
  cosmological dark matter}},
  \href{https://doi.org/10.1103/PhysRevD.95.043541}{\emph{Phys. Rev. D}
  {\bfseries 95} (2017) 043541}
  [\href{https://arxiv.org/abs/1610.08297}{{\ttfamily 1610.08297}}].

\bibitem{Ibe:2018ffn}
M.~Ibe, M.~Yamazaki and T.T.~Yanagida, \emph{{Quintessence Axion Revisited in
  Light of Swampland Conjectures}},
  \href{https://doi.org/10.1088/1361-6382/ab5197}{\emph{Class. Quant. Grav.}
  {\bfseries 36} (2019) 235020}
  [\href{https://arxiv.org/abs/1811.04664}{{\ttfamily 1811.04664}}].

\bibitem{Mocz:2019pyf}
P.~Mocz et~al., \emph{{First star-forming structures in fuzzy cosmic
  filaments}},
  \href{https://doi.org/10.1103/PhysRevLett.123.141301}{\emph{Phys. Rev. Lett.}
  {\bfseries 123} (2019) 141301}
  [\href{https://arxiv.org/abs/1910.01653}{{\ttfamily 1910.01653}}].

\bibitem{Ferreira:2020fam}
E.G.M.~Ferreira, \emph{{Ultra-light dark matter}},
  \href{https://doi.org/10.1007/s00159-021-00135-6}{\emph{Astron. Astrophys.
  Rev.} {\bfseries 29} (2021) 7}
  [\href{https://arxiv.org/abs/2005.03254}{{\ttfamily 2005.03254}}].

\bibitem{Hui:2021tkt}
L.~Hui, \emph{{Wave Dark Matter}},
  \href{https://doi.org/10.1146/annurev-astro-120920-010024}{\emph{Ann. Rev.
  Astron. Astrophys.} {\bfseries 59} (2021) 247}
  [\href{https://arxiv.org/abs/2101.11735}{{\ttfamily 2101.11735}}].

\bibitem{Choi:2021aze}
G.~Choi, W.~Lin, L.~Visinelli and T.T.~Yanagida, \emph{{Cosmic birefringence
  and electroweak axion dark energy}},
  \href{https://doi.org/10.1103/PhysRevD.104.L101302}{\emph{Phys. Rev. D}
  {\bfseries 104} (2021) L101302}
  [\href{https://arxiv.org/abs/2106.12602}{{\ttfamily 2106.12602}}].

\bibitem{Svrcek:2006yi}
P.~Svrcek and E.~Witten, \emph{{Axions In String Theory}},
  \href{https://doi.org/10.1088/1126-6708/2006/06/051}{\emph{JHEP} {\bfseries
  06} (2006) 051} [\href{https://arxiv.org/abs/hep-th/0605206}{{\ttfamily
  hep-th/0605206}}].

\bibitem{Arvanitaki:2009fg}
A.~Arvanitaki, S.~Dimopoulos, S.~Dubovsky, N.~Kaloper and J.~March-Russell,
  \emph{{String Axiverse}},
  \href{https://doi.org/10.1103/PhysRevD.81.123530}{\emph{Phys. Rev. D}
  {\bfseries 81} (2010) 123530}
  [\href{https://arxiv.org/abs/0905.4720}{{\ttfamily 0905.4720}}].

\bibitem{Cicoli:2012sz}
M.~Cicoli, M.~Goodsell and A.~Ringwald, \emph{{The type IIB string axiverse and
  its low-energy phenomenology}},
  \href{https://doi.org/10.1007/JHEP10(2012)146}{\emph{JHEP} {\bfseries 10}
  (2012) 146} [\href{https://arxiv.org/abs/1206.0819}{{\ttfamily 1206.0819}}].

\bibitem{Visinelli:2018utg}
L.~Visinelli and S.~Vagnozzi, \emph{{Cosmological window onto the string
  axiverse and the supersymmetry breaking scale}},
  \href{https://doi.org/10.1103/PhysRevD.99.063517}{\emph{Phys. Rev. D}
  {\bfseries 99} (2019) 063517}
  [\href{https://arxiv.org/abs/1809.06382}{{\ttfamily 1809.06382}}].

\bibitem{Visinelli:2019qqu}
L.~Visinelli, S.~Vagnozzi and U.~Danielsson, \emph{{Revisiting a negative
  cosmological constant from low-redshift data}},
  \href{https://doi.org/10.3390/sym11081035}{\emph{Symmetry} {\bfseries 11}
  (2019) 1035} [\href{https://arxiv.org/abs/1907.07953}{{\ttfamily
  1907.07953}}].

\bibitem{Carone:1994aa}
C.D.~Carone and H.~Murayama, \emph{{Possible light U(1) gauge boson coupled to
  baryon number}},
  \href{https://doi.org/10.1103/PhysRevLett.74.3122}{\emph{Phys. Rev. Lett.}
  {\bfseries 74} (1995) 3122}
  [\href{https://arxiv.org/abs/hep-ph/9411256}{{\ttfamily hep-ph/9411256}}].

\bibitem{FileviezPerez:2010gw}
P.~Fileviez~Perez and M.B.~Wise, \emph{{Baryon and lepton number as local gauge
  symmetries}}, \href{https://doi.org/10.1103/PhysRevD.82.079901}{\emph{Phys.
  Rev. D} {\bfseries 82} (2010) 011901}
  [\href{https://arxiv.org/abs/1002.1754}{{\ttfamily 1002.1754}}].

\bibitem{Elor:2018twp}
G.~Elor, M.~Escudero and A.~Nelson, \emph{{Baryogenesis and Dark Matter from
  $B$ Mesons}}, \href{https://doi.org/10.1103/PhysRevD.99.035031}{\emph{Phys.
  Rev. D} {\bfseries 99} (2019) 035031}
  [\href{https://arxiv.org/abs/1810.00880}{{\ttfamily 1810.00880}}].

\bibitem{Davidson:1978pm}
A.~Davidson, \emph{{$B-L$ as the fourth color within an $\mathrm{SU}(2)_L
  \times \mathrm{U}(1)_R \times \mathrm{U}(1)$ model}},
  \href{https://doi.org/10.1103/PhysRevD.20.776}{\emph{Phys. Rev. D} {\bfseries
  20} (1979) 776}.

\bibitem{Mohapatra:1980qe}
R.N.~Mohapatra and R.E.~Marshak, \emph{{Local B-L Symmetry of Electroweak
  Interactions, Majorana Neutrinos and Neutron Oscillations}},
  \href{https://doi.org/10.1103/PhysRevLett.44.1316}{\emph{Phys. Rev. Lett.}
  {\bfseries 44} (1980) 1316}.

\bibitem{Davidson:1987mh}
A.~Davidson and K.C.~Wali, \emph{{Universal Seesaw Mechanism?}},
  \href{https://doi.org/10.1103/PhysRevLett.59.393}{\emph{Phys. Rev. Lett.}
  {\bfseries 59} (1987) 393}.

\bibitem{Escudero:2018fwn}
M.~Escudero, S.J.~Witte and N.~Rius, \emph{{The dispirited case of gauged
  U(1)$_{B-L}$ dark matter}},
  \href{https://doi.org/10.1007/JHEP08(2018)190}{\emph{JHEP} {\bfseries 08}
  (2018) 190} [\href{https://arxiv.org/abs/1806.02823}{{\ttfamily
  1806.02823}}].

\bibitem{Lin:2022xbu}
W.~Lin, L.~Visinelli, D.~Xu and T.T.~Yanagida, \emph{{Neutrino astronomy as a
  probe of physics beyond the Standard Model: Decay of sub-MeV B-L gauge boson
  dark matter}}, \href{https://doi.org/10.1103/PhysRevD.106.075011}{\emph{Phys.
  Rev. D} {\bfseries 106} (2022) 075011}
  [\href{https://arxiv.org/abs/2202.04496}{{\ttfamily 2202.04496}}].

\bibitem{Foot:1990mn}
R.~Foot, \emph{{New Physics From Electric Charge Quantization?}},
  \href{https://doi.org/10.1142/S0217732391000543}{\emph{Mod. Phys. Lett. A}
  {\bfseries 6} (1991) 527}.

\bibitem{He:1991qd}
X.-G.~He, G.C.~Joshi, H.~Lew and R.R.~Volkas, \emph{{Simplest Z-prime model}},
  \href{https://doi.org/10.1103/PhysRevD.44.2118}{\emph{Phys. Rev. D}
  {\bfseries 44} (1991) 2118}.

\bibitem{Escudero:2019gzq}
M.~Escudero, D.~Hooper, G.~Krnjaic and M.~Pierre, \emph{{Cosmology with A Very
  Light L$_{\mu}$ \ensuremath{-} L$_{\tau}$ Gauge Boson}},
  \href{https://doi.org/10.1007/JHEP03(2019)071}{\emph{JHEP} {\bfseries 03}
  (2019) 071} [\href{https://arxiv.org/abs/1901.02010}{{\ttfamily
  1901.02010}}].

\bibitem{Izaguirre:2014cza}
E.~Izaguirre, G.~Krnjaic and M.~Pospelov, \emph{{Probing New Physics with
  Underground Accelerators and Radioactive Sources}},
  \href{https://doi.org/10.1016/j.physletb.2014.11.037}{\emph{Phys. Lett. B}
  {\bfseries 740} (2015) 61} [\href{https://arxiv.org/abs/1405.4864}{{\ttfamily
  1405.4864}}].

\bibitem{Pospelov:2017kep}
M.~Pospelov and Y.-D.~Tsai, \emph{{Light scalars and dark photons in Borexino
  and LSND experiments}},
  \href{https://doi.org/10.1016/j.physletb.2018.08.053}{\emph{Phys. Lett. B}
  {\bfseries 785} (2018) 288}
  [\href{https://arxiv.org/abs/1706.00424}{{\ttfamily 1706.00424}}].

\bibitem{Blinov:2018vgc}
N.~Blinov, S.A.R.~Ellis and A.~Hook, \emph{{Consequences of Fine-Tuning for
  Fifth Force Searches}},
  \href{https://doi.org/10.1007/JHEP11(2018)029}{\emph{JHEP} {\bfseries 11}
  (2018) 029} [\href{https://arxiv.org/abs/1807.11508}{{\ttfamily
  1807.11508}}].

\bibitem{Sibiryakov:2020eir}
S.~Sibiryakov, P.~S\o{}rensen and T.-T.~Yu, \emph{{BBN constraints on
  universally-coupled ultralight scalar dark matter}},
  \href{https://doi.org/10.1007/JHEP12(2020)075}{\emph{JHEP} {\bfseries 12}
  (2020) 075} [\href{https://arxiv.org/abs/2006.04820}{{\ttfamily
  2006.04820}}].

\bibitem{Paliathanasis:2023ttu}
A.~Paliathanasis, \emph{{Dynamical analysis in Chameleon dark energy}},
  \href{https://arxiv.org/abs/2306.03880}{{\ttfamily 2306.03880}}.

\bibitem{Hassan:2011hr}
S.F.~Hassan and R.A.~Rosen, \emph{{Resolving the Ghost Problem in non-Linear
  Massive Gravity}},
  \href{https://doi.org/10.1103/PhysRevLett.108.041101}{\emph{Phys. Rev. Lett.}
  {\bfseries 108} (2012) 041101}
  [\href{https://arxiv.org/abs/1106.3344}{{\ttfamily 1106.3344}}].

\bibitem{Hassan:2011zd}
S.F.~Hassan and R.A.~Rosen, \emph{{Bimetric Gravity from Ghost-free Massive
  Gravity}}, \href{https://doi.org/10.1007/JHEP02(2012)126}{\emph{JHEP}
  {\bfseries 02} (2012) 126} [\href{https://arxiv.org/abs/1109.3515}{{\ttfamily
  1109.3515}}].

\bibitem{Bernus:2019rgl}
L.~Bernus, O.~Minazzoli, A.~Fienga, M.~Gastineau, J.~Laskar and P.~Deram,
  \emph{{Constraining the mass of the graviton with the planetary ephemeris
  INPOP}}, \href{https://doi.org/10.1103/PhysRevLett.123.161103}{\emph{Phys.
  Rev. Lett.} {\bfseries 123} (2019) 161103}
  [\href{https://arxiv.org/abs/1901.04307}{{\ttfamily 1901.04307}}].

\bibitem{Bernus:2020szc}
L.~Bernus, O.~Minazzoli, A.~Fienga, M.~Gastineau, J.~Laskar, P.~Deram et~al.,
  \emph{{Constraint on the Yukawa suppression of the Newtonian potential from
  the planetary ephemeris INPOP19a}},
  \href{https://doi.org/10.1103/PhysRevD.102.021501}{\emph{Phys. Rev. D}
  {\bfseries 102} (2020) 021501}
  [\href{https://arxiv.org/abs/2006.12304}{{\ttfamily 2006.12304}}].

\bibitem{Fischbach:1985tk}
E.~Fischbach, D.~Sudarsky, A.~Szafer, C.~Talmadge and S.H.~Aronson,
  \emph{{Reanalysis of the Eotvos Experiment}},
  \href{https://doi.org/10.1103/PhysRevLett.56.3}{\emph{Phys. Rev. Lett.}
  {\bfseries 56} (1986) 3}.

\bibitem{DeRujula:1986ug}
A.~De~Rujula, \emph{{On weaker forces than gravity}},
  \href{https://doi.org/10.1016/0370-2693(86)90298-4}{\emph{Phys. Lett. B}
  {\bfseries 180} (1986) 213}.

\bibitem{Talmadge:1988qz}
C.~Talmadge, J.P.~Berthias, R.W.~Hellings and E.M.~Standish, \emph{{Model
  Independent Constraints on Possible Modifications of Newtonian Gravity}},
  \href{https://doi.org/10.1103/PhysRevLett.61.1159}{\emph{Phys. Rev. Lett.}
  {\bfseries 61} (1988) 1159}.

\bibitem{Fischbach:1992fa}
E.~Fischbach and C.~Talmadge, \emph{{Six years of the fifth force}},
  \href{https://doi.org/10.1038/356207a0}{\emph{Nature} {\bfseries 356} (1992)
  207}.

\bibitem{PhysRevD.50.3614}
Y.~Su, B.R.~Heckel, E.G.~Adelberger, J.H.~Gundlach, M.~Harris, G.L.~Smith
  et~al., \emph{New tests of the universality of free fall},
  \href{https://doi.org/10.1103/PhysRevD.50.3614}{\emph{Phys. Rev. D}
  {\bfseries 50} (1994) 3614}.

\bibitem{Hoyle:2004cw}
C.D.~Hoyle, D.J.~Kapner, B.R.~Heckel, E.G.~Adelberger, J.H.~Gundlach,
  U.~Schmidt et~al., \emph{{Sub-millimeter tests of the gravitational
  inverse-square law}},
  \href{https://doi.org/10.1103/PhysRevD.70.042004}{\emph{Phys. Rev. D}
  {\bfseries 70} (2004) 042004}
  [\href{https://arxiv.org/abs/hep-ph/0405262}{{\ttfamily hep-ph/0405262}}].

\bibitem{Mota:2006fz}
D.F.~Mota and D.J.~Shaw, \emph{{Evading Equivalence Principle Violations,
  Cosmological and other Experimental Constraints in Scalar Field Theories with
  a Strong Coupling to Matter}},
  \href{https://doi.org/10.1103/PhysRevD.75.063501}{\emph{Phys. Rev. D}
  {\bfseries 75} (2007) 063501}
  [\href{https://arxiv.org/abs/hep-ph/0608078}{{\ttfamily hep-ph/0608078}}].

\bibitem{Brax:2007vm}
P.~Brax, C.~van~de Bruck, A.-C.~Davis, D.F.~Mota and D.J.~Shaw,
  \emph{{Detecting chameleons through Casimir force measurements}},
  \href{https://doi.org/10.1103/PhysRevD.76.124034}{\emph{Phys. Rev. D}
  {\bfseries 76} (2007) 124034}
  [\href{https://arxiv.org/abs/0709.2075}{{\ttfamily 0709.2075}}].

\bibitem{Schlamminger:2007ht}
S.~Schlamminger, K.Y.~Choi, T.A.~Wagner, J.H.~Gundlach and E.G.~Adelberger,
  \emph{{Test of the equivalence principle using a rotating torsion balance}},
  \href{https://doi.org/10.1103/PhysRevLett.100.041101}{\emph{Phys. Rev. Lett.}
  {\bfseries 100} (2008) 041101}
  [\href{https://arxiv.org/abs/0712.0607}{{\ttfamily 0712.0607}}].

\bibitem{Brax:2011hb}
P.~Brax and G.~Pignol, \emph{{Strongly Coupled Chameleons and the Neutronic
  Quantum Bouncer}},
  \href{https://doi.org/10.1103/PhysRevLett.107.111301}{\emph{Phys. Rev. Lett.}
  {\bfseries 107} (2011) 111301}
  [\href{https://arxiv.org/abs/1105.3420}{{\ttfamily 1105.3420}}].

\bibitem{Wagner:2012ui}
T.A.~Wagner, S.~Schlamminger, J.H.~Gundlach and E.G.~Adelberger,
  \emph{{Torsion-balance tests of the weak equivalence principle}},
  \href{https://doi.org/10.1088/0264-9381/29/18/184002}{\emph{Class. Quant.
  Grav.} {\bfseries 29} (2012) 184002}
  [\href{https://arxiv.org/abs/1207.2442}{{\ttfamily 1207.2442}}].

\bibitem{Burrage:2014oza}
C.~Burrage, E.J.~Copeland and E.A.~Hinds, \emph{{Probing Dark Energy with Atom
  Interferometry}},
  \href{https://doi.org/10.1088/1475-7516/2015/03/042}{\emph{JCAP} {\bfseries
  03} (2015) 042} [\href{https://arxiv.org/abs/1408.1409}{{\ttfamily
  1408.1409}}].

\bibitem{Foot:2014uba}
R.~Foot and S.~Vagnozzi, \emph{{Dissipative hidden sector dark matter}},
  \href{https://doi.org/10.1103/PhysRevD.91.023512}{\emph{Phys. Rev. D}
  {\bfseries 91} (2015) 023512}
  [\href{https://arxiv.org/abs/1409.7174}{{\ttfamily 1409.7174}}].

\bibitem{Foot:2014osa}
R.~Foot and S.~Vagnozzi, \emph{{Diurnal modulation signal from dissipative
  hidden sector dark matter}},
  \href{https://doi.org/10.1016/j.physletb.2015.06.063}{\emph{Phys. Lett. B}
  {\bfseries 748} (2015) 61} [\href{https://arxiv.org/abs/1412.0762}{{\ttfamily
  1412.0762}}].

\bibitem{Brax:2015hma}
P.~Brax, C.~Burrage and C.~Englert, \emph{{Disformal dark energy at
  colliders}}, \href{https://doi.org/10.1103/PhysRevD.92.044036}{\emph{Phys.
  Rev. D} {\bfseries 92} (2015) 044036}
  [\href{https://arxiv.org/abs/1506.04057}{{\ttfamily 1506.04057}}].

\bibitem{Brax:2016did}
P.~Brax, C.~Burrage, C.~Englert and M.~Spannowsky, \emph{{LHC Signatures Of
  Scalar Dark Energy}},
  \href{https://doi.org/10.1103/PhysRevD.94.084054}{\emph{Phys. Rev. D}
  {\bfseries 94} (2016) 084054}
  [\href{https://arxiv.org/abs/1604.04299}{{\ttfamily 1604.04299}}].

\bibitem{TheMADMAXWorkingGroup:2016hpc}
{\scshape MADMAX Working Group} collaboration, \emph{{Dielectric Haloscopes: A
  New Way to Detect Axion Dark Matter}},
  \href{https://doi.org/10.1103/PhysRevLett.118.091801}{\emph{Phys. Rev. Lett.}
  {\bfseries 118} (2017) 091801}
  [\href{https://arxiv.org/abs/1611.05865}{{\ttfamily 1611.05865}}].

\bibitem{Perivolaropoulos:2016ucs}
L.~Perivolaropoulos, \emph{{Submillimeter spatial oscillations of
  Newton\textquoteright{}s constant: Theoretical models and laboratory tests}},
  \href{https://doi.org/10.1103/PhysRevD.95.084050}{\emph{Phys. Rev. D}
  {\bfseries 95} (2017) 084050}
  [\href{https://arxiv.org/abs/1611.07293}{{\ttfamily 1611.07293}}].

\bibitem{Burrage:2017qrf}
C.~Burrage and J.~Sakstein, \emph{{Tests of Chameleon Gravity}},
  \href{https://doi.org/10.1007/s41114-018-0011-x}{\emph{Living Rev. Rel.}
  {\bfseries 21} (2018) 1} [\href{https://arxiv.org/abs/1709.09071}{{\ttfamily
  1709.09071}}].

\bibitem{Irastorza:2018dyq}
I.G.~Irastorza and J.~Redondo, \emph{{New experimental approaches in the search
  for axion-like particles}},
  \href{https://doi.org/10.1016/j.ppnp.2018.05.003}{\emph{Prog. Part. Nucl.
  Phys.} {\bfseries 102} (2018) 89}
  [\href{https://arxiv.org/abs/1801.08127}{{\ttfamily 1801.08127}}].

\bibitem{Capolupo:2019xyd}
A.~Capolupo, I.~De~Martino, G.~Lambiase and A.~Stabile,
  \emph{{Axion\textendash{}photon mixing in quantum field theory and vacuum
  energy}}, \href{https://doi.org/10.1016/j.physletb.2019.01.056}{\emph{Phys.
  Lett. B} {\bfseries 790} (2019) 427}
  [\href{https://arxiv.org/abs/1901.10473}{{\ttfamily 1901.10473}}].

\bibitem{Perivolaropoulos:2019vkb}
L.~Perivolaropoulos and L.~Kazantzidis, \emph{{Hints of modified gravity in
  cosmos and in the lab?}},
  \href{https://doi.org/10.1142/S021827181942001X}{\emph{Int. J. Mod. Phys. D}
  {\bfseries 28} (2019) 1942001}
  [\href{https://arxiv.org/abs/1904.09462}{{\ttfamily 1904.09462}}].

\bibitem{Blanco:2019hah}
C.~Blanco, M.~Escudero, D.~Hooper and S.J.~Witte, \emph{{Z' mediated WIMPs:
  dead, dying, or soon to be detected?}},
  \href{https://doi.org/10.1088/1475-7516/2019/11/024}{\emph{JCAP} {\bfseries
  11} (2019) 024} [\href{https://arxiv.org/abs/1907.05893}{{\ttfamily
  1907.05893}}].

\bibitem{Capolupo:2019peg}
A.~Capolupo, G.~Lambiase, A.~Quaranta and S.M.~Giampaolo, \emph{{Probing axion
  mediated fermion\textendash{}fermion interaction by means of entanglement}},
  \href{https://doi.org/10.1016/j.physletb.2020.135407}{\emph{Phys. Lett. B}
  {\bfseries 804} (2020) 135407}
  [\href{https://arxiv.org/abs/1910.01533}{{\ttfamily 1910.01533}}].

\bibitem{Braine:2019fqb}
{\scshape ADMX} collaboration, \emph{{Extended Search for the Invisible Axion
  with the Axion Dark Matter Experiment}},
  \href{https://doi.org/10.1103/PhysRevLett.124.101303}{\emph{Phys. Rev. Lett.}
  {\bfseries 124} (2020) 101303}
  [\href{https://arxiv.org/abs/1910.08638}{{\ttfamily 1910.08638}}].

\bibitem{DiLuzio:2020wdo}
L.~Di~Luzio, M.~Giannotti, E.~Nardi and L.~Visinelli, \emph{{The landscape of
  QCD axion models}},
  \href{https://doi.org/10.1016/j.physrep.2020.06.002}{\emph{Phys. Rept.}
  {\bfseries 870} (2020) 1} [\href{https://arxiv.org/abs/2003.01100}{{\ttfamily
  2003.01100}}].

\bibitem{KumarPoddar:2020kdz}
T.~Kumar~Poddar, S.~Mohanty and S.~Jana, \emph{{Constraints on long range force
  from perihelion precession of planets in a gauged $L_e-L_{\mu,\tau}$
  scenario}}, \href{https://doi.org/10.1140/epjc/s10052-021-09078-9}{\emph{Eur.
  Phys. J. C} {\bfseries 81} (2021) 286}
  [\href{https://arxiv.org/abs/2002.02935}{{\ttfamily 2002.02935}}].

\bibitem{Bloch:2020uzh}
I.M.~Bloch, A.~Caputo, R.~Essig, D.~Redigolo, M.~Sholapurkar and T.~Volansky,
  \emph{{Exploring new physics with O(keV) electron recoils in direct detection
  experiments}}, \href{https://doi.org/10.1007/JHEP01(2021)178}{\emph{JHEP}
  {\bfseries 01} (2021) 178}
  [\href{https://arxiv.org/abs/2006.14521}{{\ttfamily 2006.14521}}].

\bibitem{Vagnozzi:2021quy}
S.~Vagnozzi, L.~Visinelli, P.~Brax, A.-C.~Davis and J.~Sakstein, \emph{{Direct
  detection of dark energy: The XENON1T excess and future prospects}},
  \href{https://doi.org/10.1103/PhysRevD.104.063023}{\emph{Phys. Rev. D}
  {\bfseries 104} (2021) 063023}
  [\href{https://arxiv.org/abs/2103.15834}{{\ttfamily 2103.15834}}].

\bibitem{Tsai:2021lly}
Y.-D.~Tsai, J.~Eby and M.S.~Safronova, \emph{{Direct detection of ultralight
  dark matter bound to the Sun with space quantum sensors}},
  \href{https://doi.org/10.1038/s41550-022-01833-6}{\emph{Nature Astron.}
  {\bfseries 7} (2023) 113} [\href{https://arxiv.org/abs/2112.07674}{{\ttfamily
  2112.07674}}].

\bibitem{Adams:2022pbo}
C.B.~Adams et~al., \emph{{Axion Dark Matter}},  in \emph{{Snowmass 2021}}, 3,
  2022 [\href{https://arxiv.org/abs/2203.14923}{{\ttfamily 2203.14923}}].

\bibitem{Antypas:2022asj}
D.~Antypas et~al., \emph{{New Horizons: Scalar and Vector Ultralight Dark
  Matter}},  \href{https://arxiv.org/abs/2203.14915}{{\ttfamily 2203.14915}}.

\bibitem{Lambiase:2022ucu}
G.~Lambiase, L.~Mastrototaro and L.~Visinelli, \emph{{Gravitational waves and
  neutrino oscillations in Chern-Simons axion gravity}},
  \href{https://doi.org/10.1088/1475-7516/2023/01/011}{\emph{JCAP} {\bfseries
  01} (2023) 011} [\href{https://arxiv.org/abs/2207.08067}{{\ttfamily
  2207.08067}}].

\bibitem{Poddar:2023pfj}
T.K.~Poddar, A.~Ghoshal and G.~Lambiase, \emph{{Listening to dark sirens from
  gravitational waves:\textbackslash{}it{Combined effects of fifth force,
  ultralight particle radiation, and eccentricity}}},
  \href{https://arxiv.org/abs/2302.14513}{{\ttfamily 2302.14513}}.

\bibitem{Elder:2023oar}
B.~Elder and J.~Sakstein, \emph{{Constraining the chameleon-photon coupling
  with atomic spectroscopy}},
  \href{https://arxiv.org/abs/2305.15638}{{\ttfamily 2305.15638}}.

\bibitem{Hlozek:2014lca}
R.~Hlozek, D.~Grin, D.J.E.~Marsh and P.G.~Ferreira, \emph{{A search for
  ultralight axions using precision cosmological data}},
  \href{https://doi.org/10.1103/PhysRevD.91.103512}{\emph{Phys. Rev. D}
  {\bfseries 91} (2015) 103512}
  [\href{https://arxiv.org/abs/1410.2896}{{\ttfamily 1410.2896}}].

\bibitem{Baumann:2015rya}
D.~Baumann, D.~Green, J.~Meyers and B.~Wallisch, \emph{{Phases of New Physics
  in the CMB}},
  \href{https://doi.org/10.1088/1475-7516/2016/01/007}{\emph{JCAP} {\bfseries
  01} (2016) 007} [\href{https://arxiv.org/abs/1508.06342}{{\ttfamily
  1508.06342}}].

\bibitem{DEramo:2018vss}
F.~D'Eramo, R.Z.~Ferreira, A.~Notari and J.L.~Bernal, \emph{{Hot Axions and the
  $H_0$ tension}},
  \href{https://doi.org/10.1088/1475-7516/2018/11/014}{\emph{JCAP} {\bfseries
  11} (2018) 014} [\href{https://arxiv.org/abs/1808.07430}{{\ttfamily
  1808.07430}}].

\bibitem{SimonsObservatory:2018koc}
{\scshape Simons Observatory} collaboration, \emph{{The Simons Observatory:
  Science goals and forecasts}},
  \href{https://doi.org/10.1088/1475-7516/2019/02/056}{\emph{JCAP} {\bfseries
  02} (2019) 056} [\href{https://arxiv.org/abs/1808.07445}{{\ttfamily
  1808.07445}}].

\bibitem{Poulin:2018dzj}
V.~Poulin, T.L.~Smith, D.~Grin, T.~Karwal and M.~Kamionkowski,
  \emph{{Cosmological implications of ultralight axionlike fields}},
  \href{https://doi.org/10.1103/PhysRevD.98.083525}{\emph{Phys. Rev. D}
  {\bfseries 98} (2018) 083525}
  [\href{https://arxiv.org/abs/1806.10608}{{\ttfamily 1806.10608}}].

\bibitem{Lambiase:2018lhs}
G.~Lambiase and S.~Mohanty, \emph{{Hydrogen spin oscillations in a background
  of axions and the 21-cm brightness temperature}},
  \href{https://doi.org/10.1093/mnras/staa1070}{\emph{Mon. Not. Roy. Astron.
  Soc.} {\bfseries 494} (2020) 5961}
  [\href{https://arxiv.org/abs/1804.05318}{{\ttfamily 1804.05318}}].

\bibitem{Poulin:2018cxd}
V.~Poulin, T.L.~Smith, T.~Karwal and M.~Kamionkowski, \emph{{Early Dark Energy
  Can Resolve The Hubble Tension}},
  \href{https://doi.org/10.1103/PhysRevLett.122.221301}{\emph{Phys. Rev. Lett.}
  {\bfseries 122} (2019) 221301}
  [\href{https://arxiv.org/abs/1811.04083}{{\ttfamily 1811.04083}}].

\bibitem{Vagnozzi:2019ezj}
S.~Vagnozzi, \emph{{New physics in light of the $H_0$ tension: An alternative
  view}}, \href{https://doi.org/10.1103/PhysRevD.102.023518}{\emph{Phys. Rev.
  D} {\bfseries 102} (2020) 023518}
  [\href{https://arxiv.org/abs/1907.07569}{{\ttfamily 1907.07569}}].

\bibitem{SimonsObservatory:2019qwx}
{\scshape Simons Observatory} collaboration, \emph{{The Simons Observatory:
  Astro2020 Decadal Project Whitepaper}}, {\emph{Bull. Am. Astron. Soc.}
  {\bfseries 51} (2019) 147}
  [\href{https://arxiv.org/abs/1907.08284}{{\ttfamily 1907.08284}}].

\bibitem{Vagnozzi:2019kvw}
S.~Vagnozzi, L.~Visinelli, O.~Mena and D.F.~Mota, \emph{{Do we have any hope of
  detecting scattering between dark energy and baryons through cosmology?}},
  \href{https://doi.org/10.1093/mnras/staa311}{\emph{Mon. Not. Roy. Astron.
  Soc.} {\bfseries 493} (2020) 1139}
  [\href{https://arxiv.org/abs/1911.12374}{{\ttfamily 1911.12374}}].

\bibitem{Green:2019glg}
D.~Green et~al., \emph{{Messengers from the Early Universe: Cosmic Neutrinos
  and Other Light Relics}}, {\emph{Bull. Am. Astron. Soc.} {\bfseries 51}
  (2019) 159} [\href{https://arxiv.org/abs/1903.04763}{{\ttfamily
  1903.04763}}].

\bibitem{EscuderoAbenza:2020cmq}
M.~Escudero~Abenza, \emph{{Precision early universe thermodynamics made simple:
  $N_{\rm eff}$ and neutrino decoupling in the Standard Model and beyond}},
  \href{https://doi.org/10.1088/1475-7516/2020/05/048}{\emph{JCAP} {\bfseries
  05} (2020) 048} [\href{https://arxiv.org/abs/2001.04466}{{\ttfamily
  2001.04466}}].

\bibitem{Odintsov:2020iui}
S.D.~Odintsov and V.K.~Oikonomou, \emph{{Aspects of Axion $F(R)$ Gravity}},
  \href{https://doi.org/10.1209/0295-5075/129/40001}{\emph{EPL} {\bfseries 129}
  (2020) 40001} [\href{https://arxiv.org/abs/2003.06671}{{\ttfamily
  2003.06671}}].

\bibitem{Rogers:2020ltq}
K.K.~Rogers and H.V.~Peiris, \emph{{Strong Bound on Canonical Ultralight Axion
  Dark Matter from the Lyman-Alpha Forest}},
  \href{https://doi.org/10.1103/PhysRevLett.126.071302}{\emph{Phys. Rev. Lett.}
  {\bfseries 126} (2021) 071302}
  [\href{https://arxiv.org/abs/2007.12705}{{\ttfamily 2007.12705}}].

\bibitem{Giare:2020vzo}
W.~Giar\`e, E.~Di~Valentino, A.~Melchiorri and O.~Mena, \emph{{New cosmological
  bounds on hot relics: axions and neutrinos}},
  \href{https://doi.org/10.1093/mnras/stab1442}{\emph{Mon. Not. Roy. Astron.
  Soc.} {\bfseries 505} (2021) 2703}
  [\href{https://arxiv.org/abs/2011.14704}{{\ttfamily 2011.14704}}].

\bibitem{Esteban:2021ozz}
I.~Esteban and J.~Salvado, \emph{{Long Range Interactions in Cosmology:
  Implications for Neutrinos}},
  \href{https://doi.org/10.1088/1475-7516/2021/05/036}{\emph{JCAP} {\bfseries
  05} (2021) 036} [\href{https://arxiv.org/abs/2101.05804}{{\ttfamily
  2101.05804}}].

\bibitem{DiValentino:2021izs}
E.~Di~Valentino, O.~Mena, S.~Pan, L.~Visinelli, W.~Yang, A.~Melchiorri et~al.,
  \emph{{In the realm of the Hubble tension\textemdash{}a review of
  solutions}}, \href{https://doi.org/10.1088/1361-6382/ac086d}{\emph{Class.
  Quant. Grav.} {\bfseries 38} (2021) 153001}
  [\href{https://arxiv.org/abs/2103.01183}{{\ttfamily 2103.01183}}].

\bibitem{Vagnozzi:2021gjh}
S.~Vagnozzi, \emph{{Consistency tests of \ensuremath{\Lambda}CDM from the early
  integrated Sachs-Wolfe effect: Implications for early-time new physics and
  the Hubble tension}},
  \href{https://doi.org/10.1103/PhysRevD.104.063524}{\emph{Phys. Rev. D}
  {\bfseries 104} (2021) 063524}
  [\href{https://arxiv.org/abs/2105.10425}{{\ttfamily 2105.10425}}].

\bibitem{Perivolaropoulos:2021jda}
L.~Perivolaropoulos and F.~Skara, \emph{{Challenges for
  \ensuremath{\Lambda}CDM: An update}},
  \href{https://doi.org/10.1016/j.newar.2022.101659}{\emph{New Astron. Rev.}
  {\bfseries 95} (2022) 101659}
  [\href{https://arxiv.org/abs/2105.05208}{{\ttfamily 2105.05208}}].

\bibitem{Xu:2021rwg}
W.L.~Xu, J.B.~Mu\~noz and C.~Dvorkin, \emph{{Cosmological constraints on light
  but massive relics}},
  \href{https://doi.org/10.1103/PhysRevD.105.095029}{\emph{Phys. Rev. D}
  {\bfseries 105} (2022) 095029}
  [\href{https://arxiv.org/abs/2107.09664}{{\ttfamily 2107.09664}}].

\bibitem{Ferlito:2022mok}
F.~Ferlito, S.~Vagnozzi, D.F.~Mota and M.~Baldi, \emph{{Cosmological direct
  detection of dark energy: Non-linear structure formation signatures of dark
  energy scattering with visible matter}},
  \href{https://doi.org/10.1093/mnras/stac649}{\emph{Mon. Not. Roy. Astron.
  Soc.} {\bfseries 512} (2022) 1885}
  [\href{https://arxiv.org/abs/2201.04528}{{\ttfamily 2201.04528}}].

\bibitem{Banerjee:2022era}
A.~Banerjee, S.~Das, A.~Maharana and R.~Kumar~Sharma, \emph{{Signatures of
  Light Massive Relics on non-linear structure formation}},
  \href{https://doi.org/10.1093/mnras/stac2128}{\emph{Mon. Not. Roy. Astron.
  Soc.} {\bfseries 516} (2022) 2038}
  [\href{https://arxiv.org/abs/2202.09840}{{\ttfamily 2202.09840}}].

\bibitem{DEramo:2022nvb}
F.~D'Eramo, E.~Di~Valentino, W.~Giar\`e, F.~Hajkarim, A.~Melchiorri, O.~Mena
  et~al., \emph{{Cosmological bound on the QCD axion mass, redux}},
  \href{https://doi.org/10.1088/1475-7516/2022/09/022}{\emph{JCAP} {\bfseries
  09} (2022) 022} [\href{https://arxiv.org/abs/2205.07849}{{\ttfamily
  2205.07849}}].

\bibitem{Bolton:2022hpt}
J.S.~Bolton, A.~Caputo, H.~Liu and M.~Viel, \emph{{Comparison of Low-Redshift
  Lyman-\ensuremath{\alpha} Forest Observations to Hydrodynamical Simulations
  with Dark Photon Dark Matter}},
  \href{https://doi.org/10.1103/PhysRevLett.129.211102}{\emph{Phys. Rev. Lett.}
  {\bfseries 129} (2022) 211102}
  [\href{https://arxiv.org/abs/2206.13520}{{\ttfamily 2206.13520}}].

\bibitem{Mazde:2022sdx}
K.~Mazde and L.~Visinelli, \emph{{The interplay between the dark matter axion
  and primordial black holes}},
  \href{https://doi.org/10.1088/1475-7516/2023/01/021}{\emph{JCAP} {\bfseries
  01} (2023) 021} [\href{https://arxiv.org/abs/2209.14307}{{\ttfamily
  2209.14307}}].

\bibitem{DiValentino:2022edq}
E.~Di~Valentino, S.~Gariazzo, W.~Giar\`e, A.~Melchiorri, O.~Mena and F.~Renzi,
  \emph{{Novel model-marginalized cosmological bound on the QCD axion mass}},
  \href{https://doi.org/10.1103/PhysRevD.107.103528}{\emph{Phys. Rev. D}
  {\bfseries 107} (2023) 103528}
  [\href{https://arxiv.org/abs/2212.11926}{{\ttfamily 2212.11926}}].

\bibitem{Rogers:2023ezo}
K.K.~Rogers, R.~Hlo\v{z}ek, A.~Lagu\"e, M.M.~Ivanov, O.H.E.~Philcox, G.~Cabass
  et~al., \emph{{Ultra-light axions and the S $_{8}$ tension: joint constraints
  from the cosmic microwave background and galaxy clustering}},
  \href{https://doi.org/10.1088/1475-7516/2023/06/023}{\emph{JCAP} {\bfseries
  06} (2023) 023} [\href{https://arxiv.org/abs/2301.08361}{{\ttfamily
  2301.08361}}].

\bibitem{Oikonomou:2023kqi}
V.K.~Oikonomou, F.P.~Fronimos, P.~Tsyba and O.~Razina, \emph{{Kinetic axion
  dark matter in string corrected f(R) gravity}},
  \href{https://doi.org/10.1016/j.dark.2023.101186}{\emph{Phys. Dark Univ.}
  {\bfseries 40} (2023) 101186}
  [\href{https://arxiv.org/abs/2302.07147}{{\ttfamily 2302.07147}}].

\bibitem{Vanzan:2023gui}
E.~Vanzan, A.~Raccanelli and N.~Bartolo, \emph{{Dark ages, a window on the dark
  sector. Hunting for ultra-light axions}},
  \href{https://arxiv.org/abs/2306.09252}{{\ttfamily 2306.09252}}.

\bibitem{Jain:2012tn}
B.~Jain, V.~Vikram and J.~Sakstein, \emph{{Astrophysical Tests of Modified
  Gravity: Constraints from Distance Indicators in the Nearby Universe}},
  \href{https://doi.org/10.1088/0004-637X/779/1/39}{\emph{Astrophys. J.}
  {\bfseries 779} (2013) 39} [\href{https://arxiv.org/abs/1204.6044}{{\ttfamily
  1204.6044}}].

\bibitem{Giannotti:2015kwo}
M.~Giannotti, I.~Irastorza, J.~Redondo and A.~Ringwald, \emph{{Cool WISPs for
  stellar cooling excesses}},
  \href{https://doi.org/10.1088/1475-7516/2016/05/057}{\emph{JCAP} {\bfseries
  05} (2016) 057} [\href{https://arxiv.org/abs/1512.08108}{{\ttfamily
  1512.08108}}].

\bibitem{Foot:2016wvj}
R.~Foot and S.~Vagnozzi, \emph{{Solving the small-scale structure puzzles with
  dissipative dark matter}},
  \href{https://doi.org/10.1088/1475-7516/2016/07/013}{\emph{JCAP} {\bfseries
  07} (2016) 013} [\href{https://arxiv.org/abs/1602.02467}{{\ttfamily
  1602.02467}}].

\bibitem{Caputo:2017zqh}
A.~Caputo, J.~Zavala and D.~Blas, \emph{{Binary pulsars as probes of a Galactic
  dark matter disk}},
  \href{https://doi.org/10.1016/j.dark.2017.10.005}{\emph{Phys. Dark Univ.}
  {\bfseries 19} (2018) 1} [\href{https://arxiv.org/abs/1709.03991}{{\ttfamily
  1709.03991}}].

\bibitem{Stott:2018opm}
M.J.~Stott and D.J.E.~Marsh, \emph{{Black hole spin constraints on the mass
  spectrum and number of axionlike fields}},
  \href{https://doi.org/10.1103/PhysRevD.98.083006}{\emph{Phys. Rev. D}
  {\bfseries 98} (2018) 083006}
  [\href{https://arxiv.org/abs/1805.02016}{{\ttfamily 1805.02016}}].

\bibitem{Roy:2019esk}
R.~Roy and U.A.~Yajnik, \emph{{Evolution of black hole shadow in the presence
  of ultralight bosons}},
  \href{https://doi.org/10.1016/j.physletb.2020.135284}{\emph{Phys. Lett. B}
  {\bfseries 803} (2020) 135284}
  [\href{https://arxiv.org/abs/1906.03190}{{\ttfamily 1906.03190}}].

\bibitem{Davoudiasl:2019nlo}
H.~Davoudiasl and P.B.~Denton, \emph{{Ultralight Boson Dark Matter and Event
  Horizon Telescope Observations of M87*}},
  \href{https://doi.org/10.1103/PhysRevLett.123.021102}{\emph{Phys. Rev. Lett.}
  {\bfseries 123} (2019) 021102}
  [\href{https://arxiv.org/abs/1904.09242}{{\ttfamily 1904.09242}}].

\bibitem{Nojiri:2019riz}
S.~Nojiri, S.D.~Odintsov and V.K.~Oikonomou, \emph{{$F(R)$ Gravity with an
  Axion-like Particle: Dynamics, Gravity Waves, Late and Early-time
  Phenomenology}},
  \href{https://doi.org/10.1016/j.aop.2020.168186}{\emph{Annals Phys.}
  {\bfseries 418} (2020) 168186}
  [\href{https://arxiv.org/abs/1907.01625}{{\ttfamily 1907.01625}}].

\bibitem{Nojiri:2019nar}
S.~Nojiri, S.D.~Odintsov, V.K.~Oikonomou and A.A.~Popov, \emph{{Propagation of
  Gravitational Waves in Chern-Simons Axion Einstein Gravity}},
  \href{https://doi.org/10.1103/PhysRevD.100.084009}{\emph{Phys. Rev. D}
  {\bfseries 100} (2019) 084009}
  [\href{https://arxiv.org/abs/1909.01324}{{\ttfamily 1909.01324}}].

\bibitem{Kumar:2020hgm}
R.~Kumar, S.G.~Ghosh and A.~Wang, \emph{{Gravitational deflection of light and
  shadow cast by rotating Kalb-Ramond black holes}},
  \href{https://doi.org/10.1103/PhysRevD.101.104001}{\emph{Phys. Rev. D}
  {\bfseries 101} (2020) 104001}
  [\href{https://arxiv.org/abs/2001.00460}{{\ttfamily 2001.00460}}].

\bibitem{Nojiri:2020pqr}
S.~Nojiri, S.D.~Odintsov, V.K.~Oikonomou and A.A.~Popov, \emph{{Propagation of
  gravitational waves in Chern\textendash{}Simons axion $F(R)$ gravity}},
  \href{https://doi.org/10.1016/j.dark.2020.100514}{\emph{Phys. Dark Univ.}
  {\bfseries 28} (2020) 100514}
  [\href{https://arxiv.org/abs/2002.10402}{{\ttfamily 2002.10402}}].

\bibitem{Creci:2020mfg}
G.~Creci, S.~Vandoren and H.~Witek, \emph{{Evolution of black hole shadows from
  superradiance}},
  \href{https://doi.org/10.1103/PhysRevD.101.124051}{\emph{Phys. Rev. D}
  {\bfseries 101} (2020) 124051}
  [\href{https://arxiv.org/abs/2004.05178}{{\ttfamily 2004.05178}}].

\bibitem{Khodadi:2020jij}
M.~Khodadi, A.~Allahyari, S.~Vagnozzi and D.F.~Mota, \emph{{Black holes with
  scalar hair in light of the Event Horizon Telescope}},
  \href{https://doi.org/10.1088/1475-7516/2020/09/026}{\emph{JCAP} {\bfseries
  09} (2020) 026} [\href{https://arxiv.org/abs/2005.05992}{{\ttfamily
  2005.05992}}].

\bibitem{Croon:2020oga}
D.~Croon, S.D.~McDermott and J.~Sakstein, \emph{{New physics and the black hole
  mass gap}}, \href{https://doi.org/10.1103/PhysRevD.102.115024}{\emph{Phys.
  Rev. D} {\bfseries 102} (2020) 115024}
  [\href{https://arxiv.org/abs/2007.07889}{{\ttfamily 2007.07889}}].

\bibitem{Stott:2020gjj}
M.J.~Stott, \emph{{Ultralight Bosonic Field Mass Bounds from Astrophysical
  Black Hole Spin}},  \href{https://arxiv.org/abs/2009.07206}{{\ttfamily
  2009.07206}}.

\bibitem{Desmond:2020nde}
H.~Desmond, J.~Sakstein and B.~Jain, \emph{{Five percent measurement of the
  gravitational constant in the Large Magellanic Cloud}},
  \href{https://doi.org/10.1103/PhysRevD.103.024028}{\emph{Phys. Rev. D}
  {\bfseries 103} (2021) 024028}
  [\href{https://arxiv.org/abs/2012.05028}{{\ttfamily 2012.05028}}].

\bibitem{Caputo:2021efm}
A.~Caputo, S.J.~Witte, D.~Blas and P.~Pani, \emph{{Electromagnetic signatures
  of dark photon superradiance}},
  \href{https://doi.org/10.1103/PhysRevD.104.043006}{\emph{Phys. Rev. D}
  {\bfseries 104} (2021) 043006}
  [\href{https://arxiv.org/abs/2102.11280}{{\ttfamily 2102.11280}}].

\bibitem{Khodadi:2021owg}
M.~Khodadi, \emph{{Black Hole Superradiance in the Presence of Lorentz Symmetry
  Violation}}, \href{https://doi.org/10.1103/PhysRevD.103.064051}{\emph{Phys.
  Rev. D} {\bfseries 103} (2021) 064051}
  [\href{https://arxiv.org/abs/2103.03611}{{\ttfamily 2103.03611}}].

\bibitem{Mehta:2021pwf}
V.M.~Mehta, M.~Demirtas, C.~Long, D.J.E.~Marsh, L.~McAllister and M.J.~Stott,
  \emph{{Superradiance in string theory}},
  \href{https://doi.org/10.1088/1475-7516/2021/07/033}{\emph{JCAP} {\bfseries
  07} (2021) 033} [\href{https://arxiv.org/abs/2103.06812}{{\ttfamily
  2103.06812}}].

\bibitem{Caputo:2021eaa}
A.~Caputo, A.J.~Millar, C.A.J.~O'Hare and E.~Vitagliano, \emph{{Dark photon
  limits: A handbook}},
  \href{https://doi.org/10.1103/PhysRevD.104.095029}{\emph{Phys. Rev. D}
  {\bfseries 104} (2021) 095029}
  [\href{https://arxiv.org/abs/2105.04565}{{\ttfamily 2105.04565}}].

\bibitem{Tsai:2021irw}
Y.-D.~Tsai, Y.~Wu, S.~Vagnozzi and L.~Visinelli, \emph{{Novel constraints on
  fifth forces and ultralight dark sector with asteroidal data}},
  \href{https://doi.org/10.1088/1475-7516/2023/04/031}{\emph{JCAP} {\bfseries
  04} (2023) 031} [\href{https://arxiv.org/abs/2107.04038}{{\ttfamily
  2107.04038}}].

\bibitem{Atamurotov:2021cgh}
F.~Atamurotov, K.~Jusufi, M.~Jamil, A.~Abdujabbarov and M.~Azreg-A\"\i{}nou,
  \emph{{Axion-plasmon or magnetized plasma effect on an observable shadow and
  gravitational lensing of a Schwarzschild black hole}},
  \href{https://doi.org/10.1103/PhysRevD.104.064053}{\emph{Phys. Rev. D}
  {\bfseries 104} (2021) 064053}
  [\href{https://arxiv.org/abs/2109.08150}{{\ttfamily 2109.08150}}].

\bibitem{Calza:2021czr}
M.~Calz\`a, J.~March-Russell and J.a.G.~Rosa, \emph{{Evaporating primordial
  black holes, the string axiverse, and hot dark radiation}},
  \href{https://arxiv.org/abs/2110.13602}{{\ttfamily 2110.13602}}.

\bibitem{Poddar:2021ose}
T.K.~Poddar, \emph{{Constraints on ultralight axions, vector gauge bosons, and
  unparticles from geodetic and frame-dragging measurements}},
  \href{https://doi.org/10.1140/epjc/s10052-022-10956-z}{\emph{Eur. Phys. J. C}
  {\bfseries 82} (2022) 982}
  [\href{https://arxiv.org/abs/2111.05632}{{\ttfamily 2111.05632}}].

\bibitem{Sakstein:2022tby}
J.~Sakstein, D.~Croon and S.D.~McDermott, \emph{{Axion instability
  supernovae}}, \href{https://doi.org/10.1103/PhysRevD.105.095038}{\emph{Phys.
  Rev. D} {\bfseries 105} (2022) 095038}
  [\href{https://arxiv.org/abs/2203.06160}{{\ttfamily 2203.06160}}].

\bibitem{Dekens:2022gha}
W.~Dekens, J.~de~Vries and S.~Shain, \emph{{CP-violating axion interactions in
  effective field theory}},
  \href{https://doi.org/10.1007/JHEP07(2022)014}{\emph{JHEP} {\bfseries 07}
  (2022) 014} [\href{https://arxiv.org/abs/2203.11230}{{\ttfamily
  2203.11230}}].

\bibitem{Vagnozzi:2022moj}
S.~Vagnozzi et~al., \emph{{Horizon-scale tests of gravity theories and
  fundamental physics from the Event Horizon Telescope image of Sagittarius
  A}}, \href{https://doi.org/10.1088/1361-6382/acd97b}{\emph{Class. Quant.
  Grav.} {\bfseries 40} (2023) 165007}
  [\href{https://arxiv.org/abs/2205.07787}{{\ttfamily 2205.07787}}].

\bibitem{Cannizzaro:2022xyw}
E.~Cannizzaro, L.~Sberna, A.~Caputo and P.~Pani, \emph{{Dark photon
  superradiance quenched by dark matter}},
  \href{https://doi.org/10.1103/PhysRevD.106.083019}{\emph{Phys. Rev. D}
  {\bfseries 106} (2022) 083019}
  [\href{https://arxiv.org/abs/2206.12367}{{\ttfamily 2206.12367}}].

\bibitem{Calza:2022ioe}
M.~Calz\'a, \emph{{Evaporation of a Kerr-black-bounce by emission of scalar
  particles}}, \href{https://doi.org/10.1103/PhysRevD.107.044067}{\emph{Phys.
  Rev. D} {\bfseries 107} (2023) 044067}
  [\href{https://arxiv.org/abs/2207.10467}{{\ttfamily 2207.10467}}].

\bibitem{Saha:2022hcd}
A.K.~Saha, P.~Parashari, T.N.~Maity, A.~Dubey, S.~Bouri and R.~Laha,
  \emph{{Bounds on ultralight bosons from the Event Horizon Telescope
  observation of Sgr A$^*$}},
  \href{https://arxiv.org/abs/2208.03530}{{\ttfamily 2208.03530}}.

\bibitem{Tsai:2022jnv}
Y.-D.~Tsai, J.~Eby, J.~Arakawa, D.~Farnocchia and M.S.~Safronova, \emph{{New
  Constraints on Dark Matter and Cosmic Neutrino Profiles through Gravity}},
  \href{https://arxiv.org/abs/2210.03749}{{\ttfamily 2210.03749}}.

\bibitem{Jha:2022tdl}
S.K.~Jha, M.~Khodadi, A.~Rahaman and A.~Sheykhi, \emph{{Superradiant energy
  extraction from rotating hairy Horndeski black holes}},
  \href{https://doi.org/10.1103/PhysRevD.107.084052}{\emph{Phys. Rev. D}
  {\bfseries 107} (2023) 084052}
  [\href{https://arxiv.org/abs/2212.13051}{{\ttfamily 2212.13051}}].

\bibitem{Unal:2023yxt}
C.~\"Unal, \emph{{Superradiance Properties of Light Black Holes and
  $10^{-12}$-$10^{21}$ eV Bosons}},
  \href{https://arxiv.org/abs/2301.08267}{{\ttfamily 2301.08267}}.

\bibitem{Oikonomou:2023bah}
V.K.~Oikonomou, \emph{{Effects of the axion through the Higgs portal on
  primordial gravitational waves during the electroweak breaking}},
  \href{https://doi.org/10.1103/PhysRevD.107.064071}{\emph{Phys. Rev. D}
  {\bfseries 107} (2023) 064071}
  [\href{https://arxiv.org/abs/2303.05889}{{\ttfamily 2303.05889}}].

\bibitem{Parbin:2023zik}
N.~Parbin, D.J.~Gogoi and U.D.~Goswami, \emph{{Weak gravitational lensing and
  shadow cast by rotating black holes in axionic Chern\textendash{}Simons
  theory}}, \href{https://doi.org/10.1016/j.dark.2023.101265}{\emph{Phys. Dark
  Univ.} {\bfseries 41} (2023) 101265}
  [\href{https://arxiv.org/abs/2305.09157}{{\ttfamily 2305.09157}}].

\bibitem{Calza:2023rjt}
M.~Calz\`a, J.a.G.~Rosa and F.~Serrano, \emph{{Primordial black hole
  superradiance and evaporation in the string axiverse}},
  \href{https://arxiv.org/abs/2306.09430}{{\ttfamily 2306.09430}}.

\bibitem{Williams:2004qba}
J.G.~Williams, S.G.~Turyshev and D.H.~Boggs, \emph{{Progress in lunar laser
  ranging tests of relativistic gravity}},
  \href{https://doi.org/10.1103/PhysRevLett.93.261101}{\emph{Phys. Rev. Lett.}
  {\bfseries 93} (2004) 261101}
  [\href{https://arxiv.org/abs/gr-qc/0411113}{{\ttfamily gr-qc/0411113}}].

\bibitem{Viswanathan:2017vob}
V.~Viswanathan, A.~Fienga, O.~Minazzoli, L.~Bernus, J.~Laskar and M.~Gastineau,
  \emph{{The new lunar ephemeris INPOP17a and its application to fundamental
  physics}}, \href{https://doi.org/10.1093/mnras/sty096}{\emph{Mon. Not. Roy.
  Astron. Soc.} {\bfseries 476} (2018) 1877}
  [\href{https://arxiv.org/abs/1710.09167}{{\ttfamily 1710.09167}}].

\bibitem{Hofmann:2018myc}
F.~Hofmann and J.~M\"uller, \emph{{Relativistic tests with lunar laser
  ranging}}, \href{https://doi.org/10.1088/1361-6382/aa8f7a}{\emph{Class.
  Quant. Grav.} {\bfseries 35} (2018) 035015}.

\bibitem{Poddar:2020exe}
T.~Kumar~Poddar, S.~Mohanty and S.~Jana, \emph{{Constraints on long range force
  from perihelion precession of planets in a gauged $L_e-L_{\mu,\tau}$
  scenario}}, \href{https://doi.org/10.1140/epjc/s10052-021-09078-9}{\emph{Eur.
  Phys. J. C} {\bfseries 81} (2021) 286}
  [\href{https://arxiv.org/abs/2002.02935}{{\ttfamily 2002.02935}}].

\bibitem{2017NSTIM.108.....V}
V.~{Viswanathan}, A.~{Fienga}, M.~{Gastineau} and J.~{Laskar}, \emph{{INPOP17a
  planetary ephemerides}},
  \href{https://doi.org/10.13140/RG.2.2.24384.43521}{\emph{Notes Scientifiques
  et Techniques de l'Institut de Mecanique Celeste} {\bfseries 108} (2017) }.

\bibitem{2019NSTIM.109.....F}
A.~{Fienga}, P.~{Deram}, V.~{Viswanathan}, A.~{Di Ruscio}, L.~{Bernus},
  D.~{Durante} et~al., \emph{{INPOP19a planetary ephemerides}}, {\emph{Notes
  Scientifiques et Techniques de l'Institut de Mecanique Celeste} {\bfseries
  109} (2019) }.

\bibitem{Fienga:2023ocw}
A.~Fienga and O.~Minazzoli, \emph{{Testing Theories of Gravity with Planetary
  Ephemerides}},  \href{https://arxiv.org/abs/2303.01821}{{\ttfamily
  2303.01821}}.

\bibitem{Mariani:2023ubf}
V.~Mariani, A.~Fienga, O.~Minazzoli, M.~Gastineau and J.~Laskar,
  \emph{{Bayesian test of the mass of the graviton with planetary
  ephemerides}}, \href{https://doi.org/10.1103/PhysRevD.108.024047}{\emph{Phys.
  Rev. D} {\bfseries 108} (2023) 024047}
  [\href{https://arxiv.org/abs/2306.07069}{{\ttfamily 2306.07069}}].

\bibitem{2022LPICo2681.2011D}
D.~{DellaGiustina}, D.R.~{Golish}, S.~{Guzewich}, M.~{Moreau}, M.C.~{Nolan},
  A.T.~{Polit} et~al., \emph{{OSIRIS-APEX: A Proposed OSIRIS-REx Extended
  Mission to Apophis}},  in \emph{LPI Contributions}, vol.~2681 of \emph{LPI
  Contributions}, p.~2011, May, 2022.

\bibitem{Moffat:2005si}
J.W.~Moffat, \emph{{Scalar-tensor-vector gravity theory}},
  \href{https://doi.org/10.1088/1475-7516/2006/03/004}{\emph{JCAP} {\bfseries
  03} (2006) 004} [\href{https://arxiv.org/abs/gr-qc/0506021}{{\ttfamily
  gr-qc/0506021}}].

\bibitem{Clifton:2011jh}
T.~Clifton, P.G.~Ferreira, A.~Padilla and C.~Skordis, \emph{{Modified Gravity
  and Cosmology}},
  \href{https://doi.org/10.1016/j.physrep.2012.01.001}{\emph{Phys. Rept.}
  {\bfseries 513} (2012) 1} [\href{https://arxiv.org/abs/1106.2476}{{\ttfamily
  1106.2476}}].

\bibitem{Giddings:1988cx}
S.B.~Giddings and A.~Strominger, \emph{{Loss of Incoherence and Determination
  of Coupling Constants in Quantum Gravity}},
  \href{https://doi.org/10.1016/0550-3213(88)90109-5}{\emph{Nucl. Phys. B}
  {\bfseries 307} (1988) 854}.

\bibitem{Coleman:1988tj}
S.R.~Coleman, \emph{{Why There Is Nothing Rather Than Something: A Theory of
  the Cosmological Constant}},
  \href{https://doi.org/10.1016/0550-3213(88)90097-1}{\emph{Nucl. Phys. B}
  {\bfseries 310} (1988) 643}.

\bibitem{Wagoner:1970vr}
R.V.~Wagoner, \emph{{Scalar tensor theory and gravitational waves}},
  \href{https://doi.org/10.1103/PhysRevD.1.3209}{\emph{Phys. Rev. D} {\bfseries
  1} (1970) 3209}.

\bibitem{Will:2014kxa}
C.M.~Will, \emph{{The Confrontation between General Relativity and
  Experiment}}, \href{https://doi.org/10.12942/lrr-2014-4}{\emph{Living Rev.
  Rel.} {\bfseries 17} (2014) 4}
  [\href{https://arxiv.org/abs/1403.7377}{{\ttfamily 1403.7377}}].

\bibitem{Adelberger:2003zx}
E.G.~Adelberger, B.R.~Heckel and A.E.~Nelson, \emph{{Tests of the gravitational
  inverse square law}},
  \href{https://doi.org/10.1146/annurev.nucl.53.041002.110503}{\emph{Ann. Rev.
  Nucl. Part. Sci.} {\bfseries 53} (2003) 77}
  [\href{https://arxiv.org/abs/hep-ph/0307284}{{\ttfamily hep-ph/0307284}}].

\bibitem{Einstein:1938yz}
A.~Einstein, L.~Infeld and B.~Hoffmann, \emph{{The Gravitational equations and
  the problem of motion}}, \href{https://doi.org/10.2307/1968714}{\emph{Annals
  Math.} {\bfseries 39} (1938) 65}.

\bibitem{Will:1993hxu}
C.M.~Will, \emph{{Theory and Experiment in Gravitational Physics}}, Cambridge
  University Press (1993),
  \href{https://doi.org/10.1017/CBO9780511564246}{10.1017/CBO9780511564246}.

\bibitem{Touboul:2017grn}
P.~Touboul et~al., \emph{{MICROSCOPE Mission: First Results of a Space Test of
  the Equivalence Principle}},
  \href{https://doi.org/10.1103/PhysRevLett.119.231101}{\emph{Phys. Rev. Lett.}
  {\bfseries 119} (2017) 231101}
  [\href{https://arxiv.org/abs/1712.01176}{{\ttfamily 1712.01176}}].

\bibitem{Fayet:2018cjy}
P.~Fayet, \emph{{MICROSCOPE limits on the strength of a new force, with
  comparisons to gravity and electromagnetism}},
  \href{https://doi.org/10.1103/PhysRevD.99.055043}{\emph{Phys. Rev. D}
  {\bfseries 99} (2019) 055043}
  [\href{https://arxiv.org/abs/1809.04991}{{\ttfamily 1809.04991}}].

\bibitem{MICROSCOPE:2022doy}
{\scshape MICROSCOPE} collaboration, \emph{{MICROSCOPE Mission: Final Results
  of the Test of the Equivalence Principle}},
  \href{https://doi.org/10.1103/PhysRevLett.129.121102}{\emph{Phys. Rev. Lett.}
  {\bfseries 129} (2022) 121102}
  [\href{https://arxiv.org/abs/2209.15487}{{\ttfamily 2209.15487}}].

\bibitem{Brito:2015oca}
R.~Brito, V.~Cardoso and P.~Pani, \emph{{Superradiance}: {New Frontiers in
  Black Hole Physics}},
  \href{https://doi.org/10.1007/978-3-319-19000-6}{\emph{Lect. Notes Phys.}
  {\bfseries 906} (2015) pp.1}
  [\href{https://arxiv.org/abs/1501.06570}{{\ttfamily 1501.06570}}].

\bibitem{Baryakhtar:2017ngi}
M.~Baryakhtar, R.~Lasenby and M.~Teo, \emph{{Black Hole Superradiance
  Signatures of Ultralight Vectors}},
  \href{https://doi.org/10.1103/PhysRevD.96.035019}{\emph{Phys. Rev. D}
  {\bfseries 96} (2017) 035019}
  [\href{https://arxiv.org/abs/1704.05081}{{\ttfamily 1704.05081}}].

\bibitem{Cardoso:2017kgn}
V.~Cardoso, P.~Pani and T.-T.~Yu, \emph{{Superradiance in rotating stars and
  pulsar-timing constraints on dark photons}},
  \href{https://doi.org/10.1103/PhysRevD.95.124056}{\emph{Phys. Rev. D}
  {\bfseries 95} (2017) 124056}
  [\href{https://arxiv.org/abs/1704.06151}{{\ttfamily 1704.06151}}].

\bibitem{2022LPICo2681.2007F}
D.~{Farnocchia} and S.R.~{Chesley}, \emph{{Apophis Trajectory, Impact Hazard,
  and Sensitivity to Spacecraft Contact}},  in \emph{Apophis T-7 Years:
  Knowledge Opportunities for the Science of Planetary Defense}, vol.~2681 of
  \emph{LPI Contributions}, p.~2007, May, 2022.

\bibitem{2018Icar..300..115B}
M.~{Brozovi{\'c}}, L.A.M.~{Benner}, J.G.~{McMichael}, J.D.~{Giorgini},
  P.~{Pravec}, P.~{Scheirich} et~al., \emph{{Goldstone and Arecibo radar
  observations of (99942) Apophis in 2012-2013}},
  \href{https://doi.org/10.1016/j.icarus.2017.08.032}{\emph{Icarus} {\bfseries
  300} (2018) 115}.

\bibitem{Iorio:2007gq}
L.~Iorio, \emph{{Constraints on the range lambda of Yukawa-like modifications
  to the Newtonian inverse-square law of gravitation from Solar System
  planetary motions}},
  \href{https://doi.org/10.1088/1126-6708/2007/10/041}{\emph{JHEP} {\bfseries
  10} (2007) 041} [\href{https://arxiv.org/abs/0708.1080}{{\ttfamily
  0708.1080}}].

\bibitem{DeMartino:2018yqf}
I.~De~Martino, R.~Lazkoz and M.~De~Laurentis, \emph{{Analysis of the Yukawa
  gravitational potential in $f(R)$ gravity I: semiclassical periastron
  advance}}, \href{https://doi.org/10.1103/PhysRevD.97.104067}{\emph{Phys. Rev.
  D} {\bfseries 97} (2018) 104067}
  [\href{https://arxiv.org/abs/1801.08135}{{\ttfamily 1801.08135}}].

\bibitem{Sun:2019ico}
B.~Sun, Z.~Cao and L.~Shao, \emph{{Constraints on fifth forces through
  perihelion precession of planets}},
  \href{https://doi.org/10.1103/PhysRevD.100.084030}{\emph{Phys. Rev. D}
  {\bfseries 100} (2019) 084030}
  [\href{https://arxiv.org/abs/1910.05666}{{\ttfamily 1910.05666}}].

\bibitem{Davis:2019ltc}
A.-C.~Davis and S.~Melville, \emph{{Novel Screening with Two Bodies: Summing
  the ladder in disformal scalar-tensor theories}},
  \href{https://doi.org/10.1088/1475-7516/2020/09/013}{\emph{JCAP} {\bfseries
  09} (2020) 013} [\href{https://arxiv.org/abs/1910.08831}{{\ttfamily
  1910.08831}}].

\bibitem{Benisty:2022txp}
D.~Benisty, \emph{{Testing modified gravity via Yukawa potential in two body
  problem: Analytical solution and observational constraints}},
  \href{https://doi.org/10.1103/PhysRevD.106.043001}{\emph{Phys. Rev. D}
  {\bfseries 106} (2022) 043001}
  [\href{https://arxiv.org/abs/2207.08235}{{\ttfamily 2207.08235}}].

\bibitem{Reynolds:2013qqa}
C.S.~Reynolds, \emph{{Measuring Black Hole Spin using X-ray Reflection
  Spectroscopy}}, \href{https://doi.org/10.1007/s11214-013-0006-6}{\emph{Space
  Sci. Rev.} {\bfseries 183} (2014) 277}
  [\href{https://arxiv.org/abs/1302.3260}{{\ttfamily 1302.3260}}].

\bibitem{Denney:2008gk}
K.D.~Denney, B.M.~Peterson, M.~Dietrich, M.~Vestergaard and M.C.~Bentz,
  \emph{{Systematic Uncertainties in Black Hole Masses Determined from Single
  Epoch Spectra}},
  \href{https://doi.org/10.1088/0004-637X/692/1/246}{\emph{Astrophys. J.}
  {\bfseries 692} (2009) 246}
  [\href{https://arxiv.org/abs/0810.3234}{{\ttfamily 0810.3234}}].

\bibitem{Shen:2013pea}
Y.~Shen, \emph{{The Mass of Quasars}}, {\emph{Bull. Astron. Soc. India}
  {\bfseries 41} (2013) 61} [\href{https://arxiv.org/abs/1302.2643}{{\ttfamily
  1302.2643}}].

\bibitem{Vagnozzi:2020quf}
S.~Vagnozzi, C.~Bambi and L.~Visinelli, \emph{{Concerns regarding the use of
  black hole shadows as standard rulers}},
  \href{https://doi.org/10.1088/1361-6382/ab7965}{\emph{Class. Quant. Grav.}
  {\bfseries 37} (2020) 087001}
  [\href{https://arxiv.org/abs/2001.02986}{{\ttfamily 2001.02986}}].

\bibitem{Arvanitaki:2014wva}
A.~Arvanitaki, M.~Baryakhtar and X.~Huang, \emph{{Discovering the QCD Axion
  with Black Holes and Gravitational Waves}},
  \href{https://doi.org/10.1103/PhysRevD.91.084011}{\emph{Phys. Rev. D}
  {\bfseries 91} (2015) 084011}
  [\href{https://arxiv.org/abs/1411.2263}{{\ttfamily 1411.2263}}].

\bibitem{Brito:2014wla}
R.~Brito, V.~Cardoso and P.~Pani, \emph{{Black holes as particle detectors:
  evolution of superradiant instabilities}},
  \href{https://doi.org/10.1088/0264-9381/32/13/134001}{\emph{Class. Quant.
  Grav.} {\bfseries 32} (2015) 134001}
  [\href{https://arxiv.org/abs/1411.0686}{{\ttfamily 1411.0686}}].

\bibitem{Roy:2021uye}
R.~Roy, S.~Vagnozzi and L.~Visinelli, \emph{{Superradiance evolution of black
  hole shadows revisited}},
  \href{https://doi.org/10.1103/PhysRevD.105.083002}{\emph{Phys. Rev. D}
  {\bfseries 105} (2022) 083002}
  [\href{https://arxiv.org/abs/2112.06932}{{\ttfamily 2112.06932}}].

\bibitem{Hui:2022sri}
L.~Hui, Y.T.A.~Law, L.~Santoni, G.~Sun, G.M.~Tomaselli and E.~Trincherini,
  \emph{{Black hole superradiance with dark matter accretion}},
  \href{https://doi.org/10.1103/PhysRevD.107.104018}{\emph{Phys. Rev. D}
  {\bfseries 107} (2023) 104018}
  [\href{https://arxiv.org/abs/2208.06408}{{\ttfamily 2208.06408}}].

\bibitem{Chen:2022nbb}
Y.~Chen, R.~Roy, S.~Vagnozzi and L.~Visinelli, \emph{{Superradiant evolution of
  the shadow and photon ring of Sgr A\ensuremath{\star}}},
  \href{https://doi.org/10.1103/PhysRevD.106.043021}{\emph{Phys. Rev. D}
  {\bfseries 106} (2022) 043021}
  [\href{https://arxiv.org/abs/2205.06238}{{\ttfamily 2205.06238}}].

\bibitem{Verma:2017ywb}
A.K.~Verma, J.-L.~Margot and A.H.~Greenberg, \emph{{Prospects of Dynamical
  Determination of General Relativity Parameter \ensuremath{\beta} and Solar
  Quadrupole Moment ${J}_{2\odot }$ with Asteroid Radar Astronomy}},
  \href{https://doi.org/10.3847/1538-4357/aa8308}{\emph{Astrophys. J.}
  {\bfseries 845} (2017) 166}
  [\href{https://arxiv.org/abs/1707.08675}{{\ttfamily 1707.08675}}].

\bibitem{SSD}
N.J.P.L.~(JPL), ``Solar system dynamics.''
  \url{https://ssd.jpl.nasa.gov//ftp/eph/planets/ioms/}, 2023.

\bibitem{2021AJ....161..105P}
R.S.~{Park}, W.M.~{Folkner}, J.G.~{Williams} and D.H.~{Boggs}, \emph{{The JPL
  Planetary and Lunar Ephemerides DE440 and DE441}},
  \href{https://doi.org/10.3847/1538-3881/abd414}{\emph{Astron. J.} {\bfseries
  161} (2021) 105}.

\bibitem{DSAC2021}
E.A.~Burt, J.D.~Prestage, R.L.~Tjoelker, D.G.~Enzer, D.~Kuang, D.W.~Murphy
  et~al., \emph{Demonstration of a trapped-ion atomic clock in space},
  \href{https://doi.org/10.1038/s41586-021-03571-7}{\emph{Nature} {\bfseries
  595} (2021) 43}.

\bibitem{Fedderke:2021kuy}
M.A.~Fedderke, P.W.~Graham and S.~Rajendran, \emph{{Asteroids for
  \ensuremath{\mu}Hz gravitational-wave detection}},
  \href{https://doi.org/10.1103/PhysRevD.105.103018}{\emph{Phys. Rev. D}
  {\bfseries 105} (2022) 103018}
  [\href{https://arxiv.org/abs/2112.11431}{{\ttfamily 2112.11431}}].

\end{thebibliography}\endgroup
\end{document}